\documentclass{aa}
\usepackage{graphicx}
\usepackage{longtable}
\usepackage{lscape}
\usepackage{amsmath}
\usepackage{lmodern}

\begin{document}

\title{Activity time series of old stars from late F to early K. V. Effect on exoplanet detectability with high-precision astrometry.}

\titlerunning{Impact of stellar activity on astrometry for old F-G-K stars V.}

\author{N. Meunier \inst{1}, A.-M. Lagrange \inst{1}, S. Borgniet \inst{2}
  }
\authorrunning{Meunier et al.}

\institute{
Univ. Grenoble Alpes, CNRS, IPAG, F-38000 Grenoble, France\\
LESIA (UMR 8109), Observatoire de Paris, PSL Research University, CNRS, UMPC, Univ. Paris Diderot, 5 Place Jules Janssen, 92195 Meudon, France \\
\email{nadege.meunier@univ-grenoble-alpes.fr}
     }

\offprints{N. Meunier}

\date{Received ; Accepted}

\abstract{ Stellar activity strongly affects and may prevent the detection of Earth-mass planets in the habitable zone of solar-type stars with radial velocity technics. Astrometry is in principle less sensitive to stellar activity because the situation is more favourable: the stellar astrometric signal is expected to be fainter than the planetary astrometric signal compared to radial velocities.  }
{We quantify the effect of stellar activity on high-precision astrometry when Earth-mass planets are searched for in the habitable zone around old main-sequence solar-type stars. }
{We used a very large set of magnetic activity synthetic time series to characterise the properties of the stellar astrometric signal. We then studied the detectability of exoplanets based on different approaches: first based on the theoretical level of false positives derived from the synthetic time series, and then with blind tests for old main-sequence F6-K4 stars. }
{The amplitude of the signal can be up to a few times the solar value depending on the assumptions made for activity level, spectral type, and spot contrast. The detection rates for 1 M$_{\rm Earth}$ planets are very good, however, with extremely  low false-positive rates in the habitable zone for stars in the F6-K4 range at 10 pc. The standard false-alarm probability using classical bootstrapping on the time series strongly overestimates the false-positive level. This  affects the detection rates.}
{We conclude that if technological challenges can be overcome and very high precision is reached, astrometry is much more suitable for detecting Earth-mass planets in the habitable zone around nearby solar-type stars than radial velocity, and  detection rates are much higher  for this range of planetary masses and periods when astrometric techniques  are used than with radial velocity techniques.}

\keywords{Astrometry    -- Stars: activity  -- Stars: solar-type -- (stars:) planetary systems} 

\maketitle

\section{Introduction}


Stellar activity affects all indirect techniques (except for those that are based on microlensing) that are used to detect exoplanets, that is,  radial velocity (RV), photometric transits, and astrometry. RV is most strongly affected because the convective blueshift is inhibited in plages \cite[][]{meunier10a,haywood16} and because several velocity fields are present at various scales, such as granulation, supergranulation, and meridional flows \cite[][]{meunier15,meunier19e,meunier19g,meunier20b}. In addition, the contrast of spots and plages also affects transits and astrometry \cite[][]{saar97,hatzes02,saar03,wright05,desort07,lagrange10b,meunier10a,boisse12,dumusque14,borgniet15,dumusque16,meunier19}.  Because of the nature of the photometric transits and their typical timescales, stellar activity mostly affects the transit characterisation of the radius and atmosphere \cite[e.g.][]{silva03,pont08,chiavassa17}  and not detectability. It is difficult, however, to reach long orbital periods, which would allow detecting Earth-like planet in the habitable zone (HZ) of solar-type stars, for example, with transits: this is one of the main goals of the  PLAnetary Transits and Oscillations of stars (PLATO) mission \cite[][]{rauer14}, which  will be launched in 2026. Only transiting planets  will be detected by PLATO, however, that is, a very small fraction of existing planetary systems. Furthermore, their mass will have to be estimated from RV follow-ups, which is expected to be difficult given the stellar activity impact.  
On the other hand, astrometry is much less affected by stellar activity than RV \cite[][]{makarov10,lagrange11}. 
GAIA is sensitive only to very massive planets (typically sub-Neptunes), which produce a signal far above the stellar activity for main-sequence stars. The main technique used so far to search for planets in the habitable zone around solar-type stars is therefore the RV technique (intermediate between photometric transits and direct imaging in terms of orbital periods)  and it allows follow-ups (microlensing techniques are not suitable for this purpose), but so far, no Earth-like planet in the habitable zone around solar-type stars has been detected because of stellar activity. 
The low impact of stellar activity on the astrometric signal compared to radial velocity is therefore one of the reasons why high-precision astrometric space missions have been proposed to detect low-mass planets around stars in our neighbourhood \cite[][]{leger15,janson18}, such as the Space Interferometry Mission (SIM) \cite[e.g.][]{shao95,svensson05,catanzarite08,makarov09}, the Nearby Earth Astrometric Telescope (NEAT) \cite[][]{malbet12,crouzier16}, or Telescope for Habitable Exoplanets and Interstellar/Intergalactic Astronomy (THEIA) \cite[][]{theia17}, although they present a huge technological challenge \cite[e.g.][]{malbet20}. 


The effect of stellar activity on the astrometric signal has first  been considered with a single-spot model \cite[][]{bastian05,reffert05} and simple spot configurations \cite[][]{eriksson07}. A statistical model of the solar signal has later been developed by \cite{catanzarite08} in the context of the SIM mission, followed by a realistic reconstruction of the solar signal over a cycle by \cite{makarov09}, with spots only, and by  \cite{makarov10} and \cite{lagrange11},  including both spots and plages. These two studies, performed for the Sun seen edge-on, agree with each other, and the latter, for example, predicts a root-mean-square (rms) of the signal over solar cycle 23 of 0.07 $\mu$as in the X direction (i.e. along the equatorial plane) and  0.05 $\mu$as in the Y direction (i.e. along the rotation axis), for a Sun at 10 pc. This is much smaller than the Earth signal (0.3 $\mu$as), and it was concluded that stellar activity was not an obstacle for detecting low-mass planets in the habitable zone around solar-type stars.  It is indeed compatible with the fact that the relative contribution of the planetary signal with respect to the stellar signal is stronger in astrometry than in radial velocities. \cite{theia17} therefore used  the solar value of 0.07 $\mu$as from \cite{lagrange11}, although they did not appear to scale it with the distance. No variability with spectral type was considered either. Finally, the effect of granulation is very small \cite[][]{svensson05} compared to the contribution from magnetic activity. \\

It is therefore necessary to study the effect of stellar activity on astrometry in more detail. First, only the Sun seen edge-on was modelled in a realistic way. The large number of simulations now available for F6-K4 stars and all stellar inclinations in \cite{meunier19}, hereafter Paper I, also allows estimating this effect more realistically.
The RV and photometric time series obtained from these simulations were analysed in \cite{meunier19b,meunier19d} and in \cite{meunier19c}. We  focus here on stellar activity, even though more effects also need to be considered for a complete model of the astrometric signal \cite[][]{sozzetti02,sozzetti05,eisner01,traub10}, such as parallaxes, the proper motion of the star, or the presence of additional more massive planets and interaction between them, which are neglected here. We take the instrumental noise into account.

Our main objective in this paper is therefore to quantify with a systematic approach the magnetic activity signal for a wide range of old main-sequence F-G-K stars and determine how this signal compares to a signal from an Earth-like planet (i.e. in terms of mass) in the habitable zone around such stars. 
 We also characterise in more detail the expected activity signal as a function of spectral type and inclination, as a reference for other applications and to determine whether astrometry may be used to characterise stellar activity. 
The outline of the paper is as follows. In Sect.~2 we describe how we model stellar activity and the planetary signal, and then explain the methods we implemented to assess exoplanet detectability. We describe the properties of stellar activity seen in astrometry in Sect.~3. The effect of stellar activity on exoplanet detectability is then analysed in detail in Sect.~4, first with a simple method based on the signal to noise ratio (S/N), then based on a theoretical false-positive  level due to stellar activity, and finally based on standard tools that are used to analyse observations, in particular, blind tests. Additional blind tests and comparisons of different techniques are presented in the appendix. We conclude in Sect.~5.

\section{Methods}

In this section, we present the stellar activity simulations, how we model the planets, and the considered instrumental configuration. The exoplanet detectability criteria used in the analysis are then discussed.

\subsection{Modelling stellar activity}

We have developed a model for the Sun, described in detail in \cite{borgniet15}, and extended it to solar-type stars in Paper I. The model produces spots, plages, and the magnetic network in a consistent way. Several observables, such as  radial velocity, photometry, chromospheric emission (converted into 
$\log R'_{HK}$ to characterise the activity level), and astrometry are produced, which provides long time series that cover one to three cycles, depending on the stars. Some input parameters depend on  spectral type, such as the activity dependence, the rotation rate, or the contrasts, while others (sizes and latitude coverage) are fixed to the solar values. The latitudinal extent is chosen as in Paper I to be similar to that of the Sun or with a maximum latitude $\theta_{\rm max}$ of the activity pattern higher than the solar one by 10$^\circ$ or 20$^\circ$. We refer to Paper I for more details.

As in Paper I, the activity levels are restricted to stars with an average $\log R'_{HK}$ below -4.5 for the most massive (F6) and below -4.85 for the less massive stars (K4), which  corresponds to the plage-dominated regime of \cite{lockwood07}. The rotation rates  were then deduced from activity-rotation relationships, increasing from a few days (F6 stars) to 30-70 days (K4). When the activity-rotation-age relationship of \cite{mamajek08} is assumed, they correspond to ages in the range of 0.5-3 Gyr for the most massive and 4-10 Gyr for the less massive stars. The plage contrasts are from  \cite{norris18}, while two laws were considered for spots: 
a lower bound, $\Delta$T$_{\rm spot1}$, equal to the solar contrast used in \cite{borgniet15},  that is, 605 K, and an upper bound law, $\Delta$T$_{\rm spot2}$, depending on T$_{\rm{eff}}$  \cite[][]{berd05},  that is, 0.75 $\times$ T$_{\rm eff}$ - 2250 K. We assumed that actual  star spots have contrasts within this range. 

A total of 22842 (11421 for each spot contrast) time series were generated, corresponding to different sets of parameters. Each of these time series was produced for ten inclinations between edge-on and pole-on configurations. The astrometry time series correspond naturally to two time series, one for the X direction (i.e. along the equatorial plane) and the other for the Y direction (i.e. along the rotation axis).

\subsection{Modelling the planet}

In this section we describe how we modelled the planetary signal and the relevant parameters. For simplicity, we considered only circular orbits, with an amplitude (in $\mu$as) $\alpha$ equal to 3 M$_{\rm pla}$ a$_{\rm pla}$ M$_{\rm star}^{-1}$ D$_{\rm star}^{-1}$  as in \cite{theia17}, where the stellar mass is given in solar mass and the planet mass in Earth mass. The planetary mass  is often chosen to be 1 M$_{\rm Earth}$, although other values are considered in  some  computations. 

The semi-major axis a$_{\rm pla}$  (in astronomical units) corresponds to the habitable zone, which we chose as in \cite{meunier19b}, depending on the spectral type \cite[][]{kasting93,jones06,zaninetti08}.  We chose a simple classical definition following \cite{kasting93} to estimate the range of distances where liquid water could be present on the surface of the planet, and taking  only luminosity effects into account: the inner side corresponds to a runaway greenhouse effect, which would imply the evaporation of all the surface water, and the outer side is the maximum distance corresponding to a temperature of 273 K in a cloud-free CO$_2$ atmosphere. This is a conservative range because additional effects could widen it \cite[see the discussion in][]{jones06}.  We focused on three orbital periods for each spectral type: the inner side of the habitable zone (PHZ$_{\rm in}$), the middle of the habitable zone (PHZ$_{\rm med}$), and the outer side of the habitable zone (PHZ$_{\rm out}$).  This corresponds to periods between 409 and 1174 days for F6 stars and between 179 and 501 days for K4 stars.

 The stellar mass (in solar mass) follows the laws  adopted in Paper I and \cite{meunier19b}. The distance D$_{\rm star}$ is usually 10 pc, unless mentioned otherwise.

A few computations were made for a star seen edge-on only, with the orbital plane of the planet assumed to be equal to the equatorial plane of the star. This is described in Appendix A.  However, for most results presented in this paper, various stellar inclinations were considered, and the planetary orbit inclination  followed two different assumptions that  are detailed in Appendix A.

\subsection{Observing parameters}

The stellar activity signal (Sect.~3) was characterised based on the full time series described in the previous section. The time series cover 3327 to 5378 days (typically 9-15 years)  depending on the simulation, with one point per night and no gap. 
All other computations that are studying the effect on exoplanet detectability were performed using the configuration proposed in the THEIA mission proposal \cite[][]{theia17}: they cover 3.5 years with 50 visits. The sampling was chosen randomly from a list of 100 random samplings, and they were also phased randomly within the cycle. For each activity simulation, one of those samplings was therefore chosen randomly, shifted randomly, and applied to this particular time series. We also added 0.199 $\mu$as noise level on each data point following the THEIA proposal. Most computations were made for a star at 10 pc: this distance affects both the activity signal and the amplitudes of the planetary signal.

No proper motion of the star was included here because we focus on stellar activity. The time series we considered are therefore 1. the activity time series alone over the full sample to characterise stellar activity in astrometry (Sect.~3), and 2. the sum of the activity time series, the planetary signal (whenever pertinent), and the Gaussian instrumental noise for  the THEIA configuration to characterise exoplanet detectability (Sect.~4).

\subsection{Principle of the analysis}

In this section, we describe how we quantify planet detectability. We use different approaches (theoretical and observational false positive levels) and criteria (frequential and temporal analysis).

\subsubsection{Signal-to-noise ratio approach}

Traditionally, $\chi^2$ probabilities have been used to assess the presence of a planetary signal in an astrometric time series \cite[][]{sozzetti02,sozzetti05,ford04,marcy05}. This is not applicable here, however, because stellar activity contributes, which could be as significant as the planetary signal.  The probabilities that indicate that a significant signal is present therefore cannot be used to indicate  the likely presence of a planet. 
A first simple approach we considered was to define a criterion based on the signal to noise ($S/N$) ratio \cite[][]{casertano99,sozzetti02,sozzetti05,eriksson07,traub10},   where the signal $S$ is equal to  $\alpha$, the amplitude of the planetary signal, and the noise $N$ is the  of the signal due to activity and instrumental noise, which can easily be computed for each  simulation (in both directions and then combining the two). A threshold on $S/N$ defines a detection limit (i.e. a value of $\alpha$ that can be converted into a planetary mass) corresponding to that threshold. This method is very easy to apply, but it has some drawbacks: the $S/N$ is computed globally, while in reality, the planetary and stellar activity signals have different frequential signatures, and the $S/N$ of a peak in a periodogram, for example, may be more relevant  than the global one. We discuss this point in Sect.~4.3. Furthermore, the method does not allow us to  estimate the  level of false positives to which this detection limit corresponds. We apply this technique in Appendix D for comparison purposes with previous works using this approach and to determine how it compares to  more sophisticated approaches.

\begin{figure}
\includegraphics{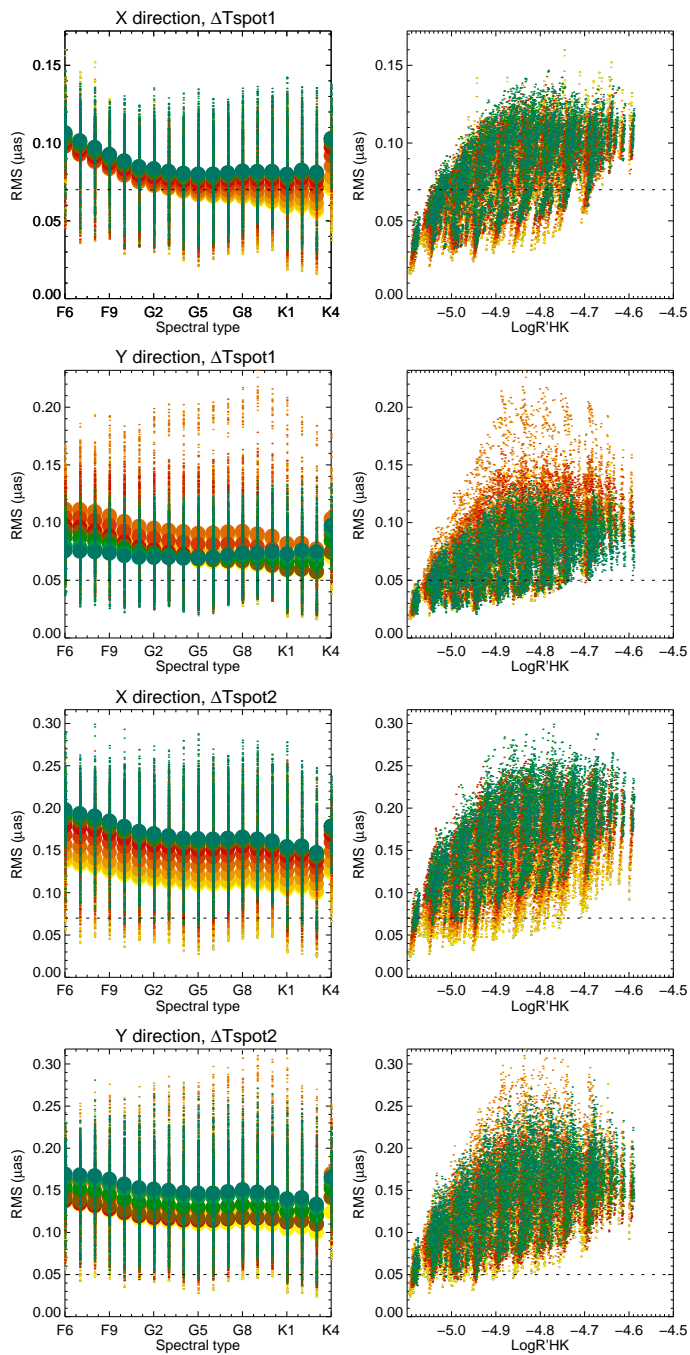}
\caption{
Rms  of the activity time series in both directions for (from upper to lower panel) the X direction and $\Delta$T$_{\rm spot1}$, the Y direction and $\Delta$T$_{\rm spot1}$, the X direction and $\Delta$T$_{\rm spot2}$, and the Y direction and $\Delta$T$_{\rm spot2}$. {\it Left column:} Rms of the astrometric signal vs. spectral type for the ten inclinations (from yellow for pole-on configurations to blue for edge-on configurations), and one simulation out of four for clarity (small dots) and binned in spectral type (circles). The horizontal dashed line is the solar value from \cite{lagrange11}. {\it Right column:} Same vs. $\log R'_{HK}$.
}
\label{act}
\end{figure}

\begin{figure*}
\includegraphics{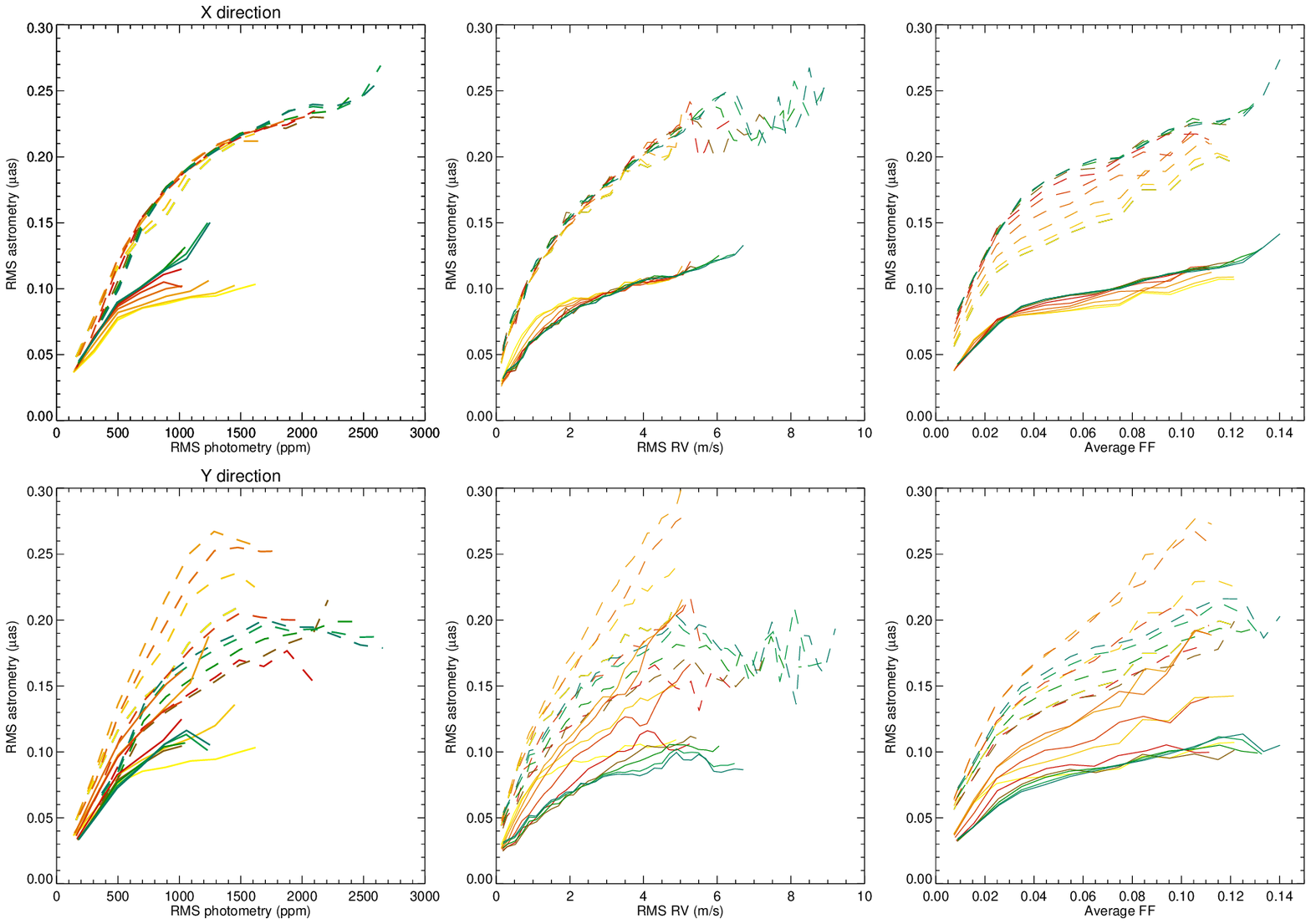}
\caption{{\it Upper panels:} Average rms of the astrometric signal in the X direction vs. rms of photometric signal (left), rms of RV signal (middle), and average filling factor (right), for the ten inclinations (colour code as in Fig.~\ref{act}) and all spectral types. Solid curves represent $\Delta$T$_{\rm spot1}$ simulations, and dashed lines represent $\Delta$T$_{\rm spot2}$ simulations. 
{\it Lower panels:} Same for the Y direction. 
}
\label{var}
\end{figure*}

\subsubsection{Detectability criteria from a theoretical point of view}

In a second step, we took the temporal variability of the time series and their frequency distribution into account. We also wish to be able to estimate a level of false positives to which the signal caused by a planet can be compared.  For this purpose, we considered two points of view, theoretical  (detailed here), and   observational  (next section 2.4.3). 
Here, we first directly estimate the level of false positives from our activity synthetic time series (without a planet). For a given criterion (either from a frequential or temporal analysis), and  assuming that our simulations correspond to a perfect knowledge of  the stellar contribution, these simulations allow us to estimate the level corresponding to 1\% of the false positives, for example. 

After adding a planet, we can compare the values based on the same criterion and estimate the percentage of cases for which the presence of the planet (of a certain mass and orbital period) leads to a signal that is stronger than this false-positive level: this gives the detection rate for this planetary mass. This approach was used, for example, by \cite{eisner01}. The two criteria and details of the computation are described in Appendix B. In this approach, we use the term "false-positive level" or "fp", which is not to be confused with the "false-alarm probability" (see next section), which we also use in this paper. In the following, the fp level always refers to a realistic level corresponding to the stellar signal alone, based on our simulations.

\subsubsection{Detectability criteria from an observational  point of view}

Our second point of view is observational. When a given star is observed, our whole set of simulations is not necessarily representative of that particular star: as a consequence, the theoretical fp level may not be correct for this particular star, and the false positive level might be desired to be estimated directly from the time series, which is in fact what is usually done. A usual way to do this is to compute a false alarm probability (FAP): we performed $N$ realisations of a bootstrap  of the time series, and computed the periodogram for each of these $N$ realisations: the maximum value in the periodogram was computed, leading to $N$ values. The 1\% (e.g.) highest values provide the 1\% FAP level.
The amplitude of the periodogram of the original time series is compared to the FAP: if it is higher than the FAP, then a significant signal is considered to be detected, with a FAP of 1\%. 
 In the following, we use the term "FAP" for this definition only.  This approach makes the strong assumption that the signal, except for the planet, follows a white-noise  distribution with the same rms as that of the original  time series. This is of course not a good assumption when the activity signal of the star  is also present because it is not white noise (and it also includes a contribution of the  planetary signal we wish to detect if present), and this will be discussed. We chose, however, to use this method because it is widely used, and we show some of its limitations in the following.  To estimate the detection rates when we used this approach, we  also performed a blind test, where the detections based on an FAP criterion were automatically compared with what was injected (planet and its properties, or no planet). The corresponding results are presented in Sect.~4.3. 

 To test this approach, we computed the FAP for a subset of simulations and compared it with the true  fp level determined above in Appendix E.
Another approach to compute detection limits from an observed time series has been proposed in  \cite{meunier12} for the radial velocity, but it can be applied to astrometric time series as well. This method, called the local power analysis (LPA), is based on the comparison between the power that would be due to a planet at a given orbital period and the maximum power in a restricted range around that orbital period. We characterise this approach and the corresponding exclusion rates in Appendix E.

\section{Properties of the activity time series}

We study the statistical properties of the activity time series as a function of spectral type, average activity level, and inclination.  Synthetic time series are also obtained for other  observables such as photometry and RV (Paper I) and are compared  to the astrometric signal.

\subsection{Dependence on spectral type and activity level}

Figure~\ref{act} shows the rms of the astrometric time series separately in the X and Y directions versus spectral type (first column) and versus the average $\log R'_{HK}$ (second column) for the different inclinations and $\Delta$T$_{\rm spot}$ values. There is a trend for an increase of the rms toward higher stellar masses and naturally toward more active stars, although there is a very large dispersion because for a given spectral type and average activity level, stars with a wide range of long-term variability are observed. The rms is  higher on average for K4 than for adjacent spectral types because quiet stars are lacking in this domain, as shown in the grid of parameters chosen in B-V and $\log R'_{HK}$ in Paper I  for these simulations (see their  Fig.~3). The rms tends to be slightly higher for the edge-on configuration than for the pole-on configuration for the X direction, while this is the opposite for the Y direction. For $\Delta$T$_{\rm spot1}$, values can be up to twice as high as the values derived for the Sun seen edge-on by \cite{lagrange11}, with a maximum of about 0.15 $\mu$as in the X direction and 0.22 $\mu$as in the Y direction. For G2 stars, the solar value is close to the average in the X direction, which is expected because in our range of parameters the Sun is also close to the average. For the Y direction, it is closer to the lower bound: the reason may be that two-thirds of the simulations are made with a higher latitudinal extent of the activity pattern (see $\theta_{\rm max}$ in Sect.~2.1) because the rms in the Y direction is sensitive to this parameter, with higher rms for high $\theta_{\rm max}$ values. For the upper boundary  of the spot contrast, however, values are higher, up to $\sim$0.3 $\mu$as, that is, comparable with the Earth signal. 
A more detailed comparison between the behaviour as a function of inclination and between the X and Y directions can be found in Appendix C, as well as the difference between the spot and plage contributions. We conclude that high inclinations are associated with a stronger variability in the X direction. The rms due to plages is of the same order of magnitude as the spot contribution when computed with $\Delta$T$_{\rm spot1}$.

\subsection{Relation between astrometry and other variables}

The average rms astrometric signal (separately in the X and Y directions) versus other variables (photometry, radial velocity, and plage filling factor) is shown in Fig.~\ref{var}. The dispersion around these averages is significant (this is due to the different spectral types and activity range covered by the simulations). A more detailed comparison between astrometry and photometry is shown in Appendix C.  We conclude that for the given conditions (inclinations and spot contrast) the relation with the variability deduced from other variables is linear, with a trend of saturation at high activity levels.

\section{Effect of activity on exoplanet detectability}

In this section, we study the effect of the stellar signal characterised in Sect.~3 on exoplanet detectability.  The detectability based on a simple estimation of the S/N is presented in Appendix D. We analyse here the results obtained with the different approaches described in Sect.~2.4, first based on theoretical fp estimated from our knowledge of stellar activity, and then using  blind tests representative of an observational approach.

\subsection{Detectability based on true false positive levels}

In this section, we consider the planetary orbit described by equations  A.4-A.9 and a distribution of the angle $\Psi$  between the orbital plane and the equatorial plane of the star described in Sect.~2.2 and Appendix A. The detections rates are first considered for a 1 M$_{\rm Earth}$, and then an iteration on the mass allows us to estimate detection limits. The orbital period takes one of the three values (inner side, middle, and outer side of the habitable zone for each spectral type).  The true fp levels derived from the whole set of synthetic time series are used, which are either constant for a given spectral type or depend on stellar variability or inclination, as described in Sect.~2.4.2 and Appendix B. 

\begin{figure}
\includegraphics{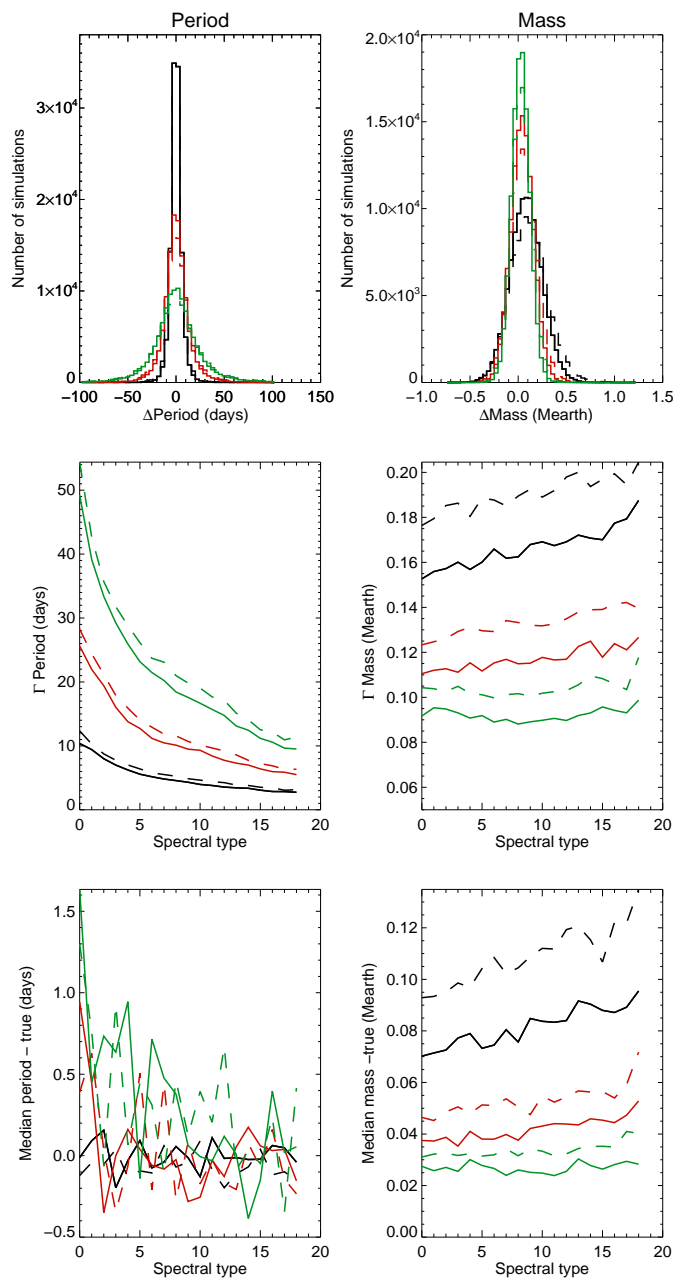}
\caption{
{\it Left panels:} Distribution of the difference between fitted period and true period for $\Delta$T$_{\rm spot1}$ (solid) and  $\Delta$T$_{\rm spot2}$ (dashed), and for the lower HZ (black), medium HZ (red), and upper HZ (green) for all simulations  (and therefore all spectral types), the distribution width for all simulations vs. spectral types, and the median  of the difference between fitted and true period vs. spectral type. 
{\it Right  panels:} Same for the mass.
}
\label{carac}
\end{figure}

\subsubsection{Periods and masses}

Figure~\ref{carac} shows the distribution  of the  departure from the true value for the orbital period (period of the peak in the periodogram) and for the planet mass for the different configurations. All spectral types are first combined together (upper panels), and the dependence of the  distribution width and of the average on spectral type  is shown in the middle and lower panels. The uncertainties on the periods are larger for the outer  side of the habitable zone than at the inner side, and lower for low-mass stars, with values of a few days and typically up to 50 days. The uncertainties on the masses are typically between 10 and 20\% (at the 1$\sigma$ level), and are higher for the inner side of the habitable zone. There is no significant bias on the period, but the mass is systematically overestimated by about 2 to 12 \% depending on the orbital period and spot contrast (which directly controls the rms of the stellar activity signal, as shown in Sect.~3). 

\subsubsection{Detection rates}

\begin{figure}
\includegraphics{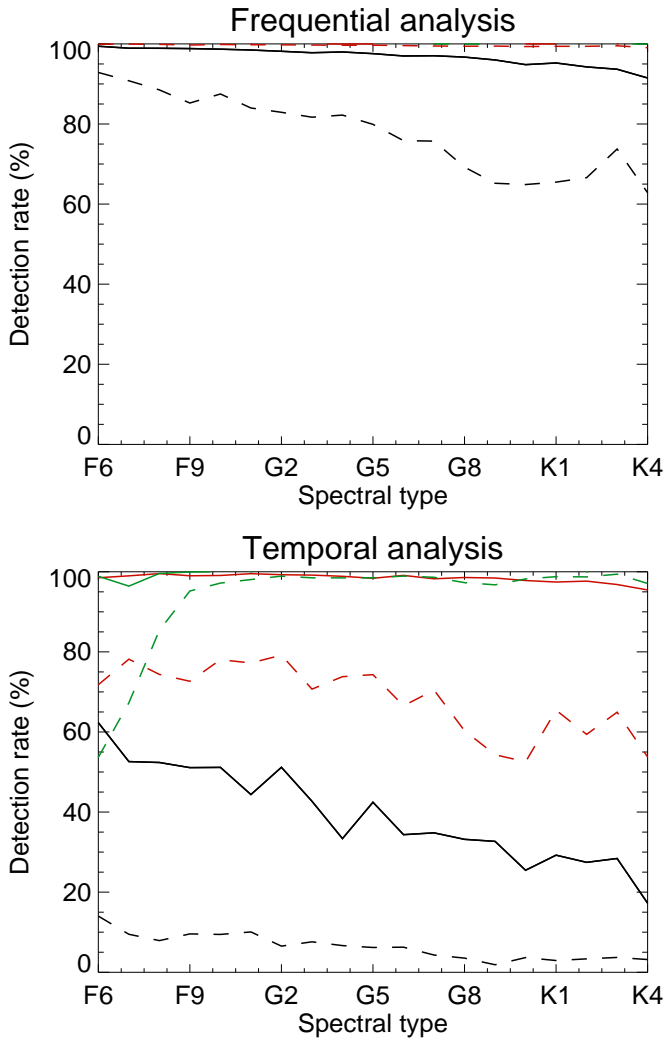}
\caption{
Detection rate vs. spectral type for 1 M$_{\rm Earth}$ planet for $\Delta$T$_{\rm spot1}$ (solid) and  $\Delta$T$_{\rm spot2}$ (dashed) for the temporal analysis (upper panel) and frequential analysis (lower panel), and for different orbital periods: lower HZ (black), medium HZ (red), and upper HZ (green).}
\label{taux_m}
\end{figure}

The same protocol allows us to  compute the detection rates based on the false positive levels described in Sect.~2.4.2. We first assumed a constant level for all simulations corresponding to a given spectral type and spot contrast (i.e. all activity levels and inclinations). The results are shown in Fig.~\ref{taux_m} for a detection criterion based on the planet mass and for a detection criterion on the power in the periodogram (i.e. temporal and frequential approach, see Sect.~2.4.2). The detection rates are excellent for the frequential approach: the detection rates are almost always at the 100\% level, except for the inner side of the habitable zone, especially with the high spot contrast (with detection rates down to 60\%).  As already illustrated in Sect.~2.4.2, the detection rates are lower for the temporal approach (for a given mass), with very low detection rates for the inner side of the habitable zone. They lie between 50\% and 100\% for the middle of the habitable zone. \cite{eisner01} and \cite{eisner02} argued that the temporal approach (i.e. a direct fit) may be more robust, but we observe here that this temporal approach leads to higher false positive levels compared to the planet signal than with the frequential approach, which is therefore more suitable. The effect of the assumptions made to compute the false positive  on the results is presented in Appendix B.3.

\subsubsection{S/N distributions}

\begin{figure}
\includegraphics{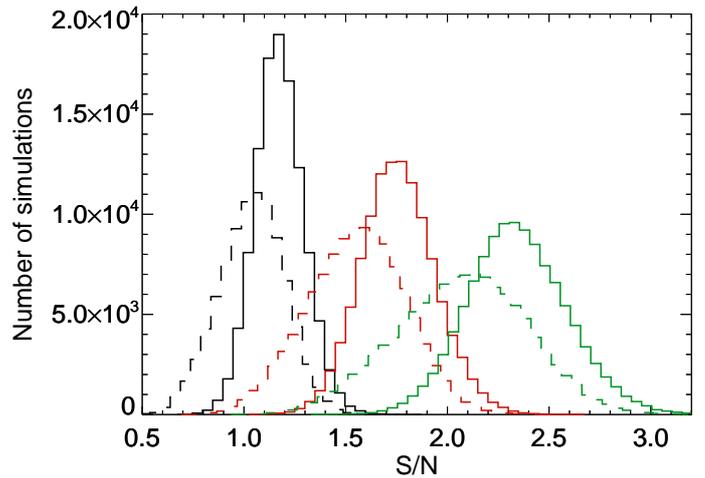}
\caption{
Distribution of the S/N for simulations above the 1\%  fp level and frequential  analysis for different configurations: $\Delta$T$_{\rm spot1}$ (solid line) and $\Delta$T$_{\rm spot2}$ (dashed line), and lower HZ (black), medium (red), and upper HZ  (green).
}
\label{test_sn}
\end{figure}

The detection rates obtained using the frequential analysis in  Fig.~\ref{taux_m} for a true level of fp of 1\% are very close to those obtained in Appendix D using the S/N$>$1 criterion. In this section, we confirm the S/N distribution corresponding to those computations. 
The S/N was computed as in Appendix D (amplitude $\alpha$ of the signal due to the 1 M$_{\rm Earth}$ planet, divided by the rms of the signal due to stellar activity and instrumental noise). 
 We show the distributions for the simulations corresponding to a planetary signal above the false positive level  and the frequential analysis (Fig.~\ref{test_sn}). 
For the low spot contrast, the S/N is in the range 0.6-1.6 for the inner side of the habitable zone (peak of the distribution around 1.2), and in the range 1.5-3.5 for the outer side  (peak around 2.4). The S/N is slightly lower for the high spot contrast. 
This shows that a relatively low global  S/N does not prevent us from obtaining good detection rates (we recall that they correspond to a 1\%  fp level). We discuss S/N issues further in Sect.~4.2.5.

\subsubsection{Detection limits}

\begin{figure}
\includegraphics{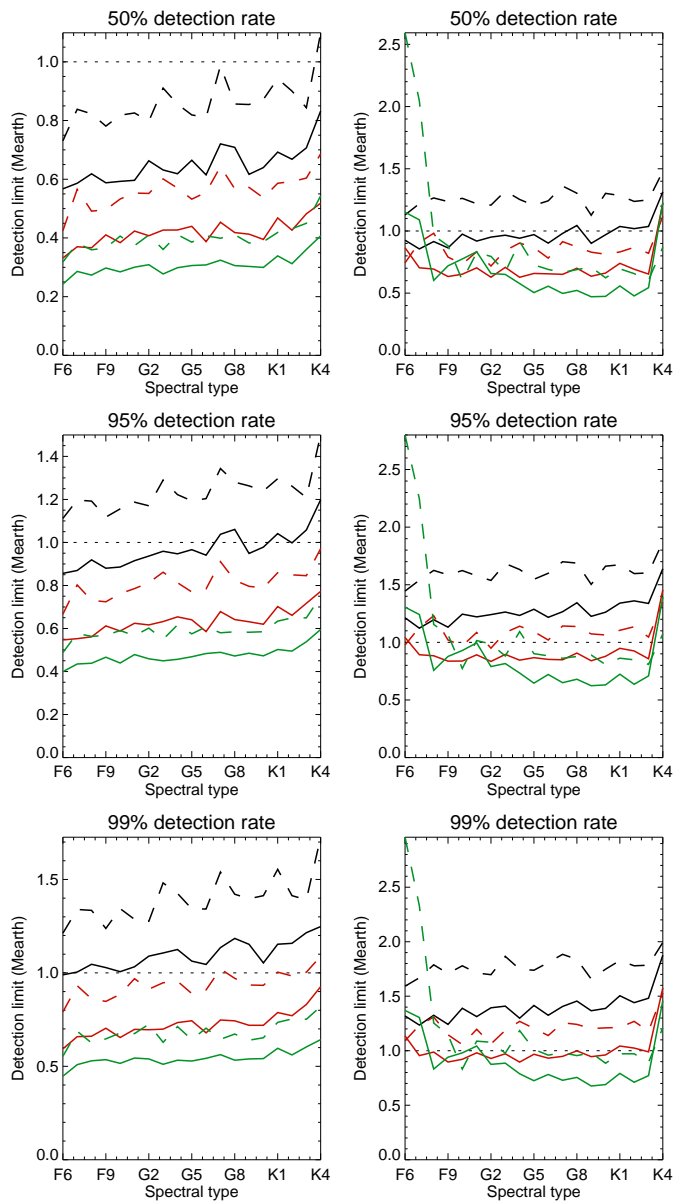}
\caption{{\it First column:} Detection limits based on a true fp level (based on power and frequential analysis) vs. spectral type for a 50\% detection rate (first panel), for a 95\% detection rate (second panel), and a for 99\% detection rate (third panel). Curves are for $\Delta$T$_{\rm spot1}$ (solid) and  $\Delta$T$_{\rm spot2}$ (dashed), and for the lower HZ (black), medium HZ (red), and upper HZ (green).
{\it Second column:} Same for fp based on a temporal  analysis (based on mass).
}
\label{fig_limdet}
\end{figure}

 Finally, we considered a more simple configuration: we focused on stars seen edge-on, with no inclination between the orbital plane and the equatorial plane, because the computations are much more time consuming and our objective is to analyse a very large set of simulations. We  first computed detection rates as in Sect.~4.1.2, but for different masses, from which we deduced a detection limit as explained in Sect.~2.4.2: this detection limit corresponds to a given detection rate (e.g. 95\%) and a given  fp level (here 1\%). The results are shown in Fig.~\ref{fig_limdet} for the temporal analysis (left panels) and for the frequential analysis (right panels) and three different levels of detection rates (50\%, 9\%, and 99\%). Again, the detection limits are higher for the temporal analysis. For a very good detection rate of 99\%, the detection limits lie around 1 M$_{\rm Earth}$ or below, except for the inner side of the habitable zone, for which they are slightly higher.
 The S/N  distributions peak in the range 0.5-2.2 for the frequential analysis and 0.6-6 for the temporal analysis, with values depending on the requested detection rate. We expect detection limits to be slightly lower for pole-on configurations.


\subsection{Detectability based on observational approach: blind tests}

In the previous section, we have computed detection rates  based on a theoretical knowledge of the false positive levels for a given set of simulations (e.g. averaged over a given spectral type): the fp level was computed over all simulations, and then applied to each of these simulations separately. In practice, the level of false positives is often determined from the observed time series itself, without any theoretical knowledge, and in particular using the FAP  defined in Sect.~2.4.3 based on a bootstrap computation. 
We therefore performed blind tests based on FAP estimation to evaluate the detection rates and false positive levels corresponding to this classical approach. Comparisons between the  FAP and  our fp levels, as well as characterisation of LPA detection limits  \cite[][]{meunier12} are performed in Appendix E.

\subsubsection{Protocol}

We performed blind tests similar to those implemented in \cite{meunier20b}. For each spectral type, we randomly selected 400 realisations of the activity simulations and randomly selected the inclination assuming uniform distribution in cos(i). The sampling was  chosen randomly as in previous sections. In statistically half the cases, no planet was injected, and a planet was injected in the remaining  simulations. The planetary parameters were chosen as follows for our reference blind test (A): 1 M$_{\rm Earth}$, period chosen randomly in the habitable zone, random phase, and angle between orbital plane and stellar equatorial plane $\Psi$ following the distribution used previously. The configuration was similar to previous tests (star at 10 pc, 0.199 $\mu$as per point, 50 points randomly chosen over 3.5 years). This was done for both spot contrasts separately.  

These time series were then automatically analysed in a simple way given the number of realisations: the FAP at the 1\% level was computed, and the highest peak in the 2-2000 day range was identified. If it is higher than the FAP, we considered it to be a detection, otherwise we considered that there is no detection for this time series (other possibilities are discussed below). If it was a detection, the fit was made as before, and the stellar inclination was assumed to be known.  The results can automatically be compared to the true parameters, which allows us to define different categories in terms of detections and false positives as described below. We note that in all our simulations, the second highest peak is almost always below the FAP, except for only 0.06\% of the cases (see Sect.~4.3.5), so that the analysis based on the highest peak above the FAP is enough.   
 The effect of the window function is not taken into account, but is expected to be small because of the random sampling: there are indeed no dominant peaks because in more than 80\% of the cases the second highest peak of the window function is within 20\% in amplitude of the highest peak (see Sect.~4.3.3 for a more detailed discussion).
In addition to this work, additional blind tests are performed to evaluate the effect of various parameters: they are listed  in Table~\ref{tab_bt}, and the averaged rates are shown and discussed in Appendix F. We conclude that the detection rates depend on distance, mass, and period, as expected, and they remain very good, except for  a higher noise level: this is therefore a critical instrumental constraint.

\begin{table}
\caption{Categories of results from blind tests}
\label{tab_bt3}
\begin{center}
\renewcommand{\footnoterule}{}  
\begin{tabular}{lll}
\hline
Detected  &  No injected  & Injected  \\
planet & injected case & planet case \\
\hline
yes  &   False positive      &  Good retrieval if good period   \\
      &                             &  False positive if incorrect period   \\
no   &   Good retrieval     &  Rejected if good period    \\
      &                             &  Missed if incorrect period   \\
\hline
\end{tabular}
\end{center}
\tablefoot{Terms used in the text for the different categories in terms of detection and false positive rates as well as the planets which are not retrieved because below the FAP. 
}
\end{table}

\subsubsection{Categories}

Depending on how the results compare to the input parameters (planet injected or not injected and its period), we identify different categories  to estimate the detection and false positive rates as in \cite{meunier20b}, adapted from \cite{dumusque17}. They are defined as follows: 1. no retrieved planet for no injection (good retrieval), 2. retrieved planet for no injection (false positive), 3. no retrieved planet although a planet was injected (characterised as a rejected planet if the period is close to the true period, and missed planet if the period is far away from the true planet), 4. retrieved planet with a good period when a planet was injected (allowing us to compute the detection rate by comparison with the number of injected planets), 5. retrieved planet with a poor period (i.e. an incorrect planet, which is a second category of false positive). 
They are summarised in Table~\ref{tab_bt3}.  The criterion for determining whether the period is close was derived from the difference distribution of the fitted period and the true period (see next paragraph) and was chosen to be 100 days.

\begin{figure}
\includegraphics{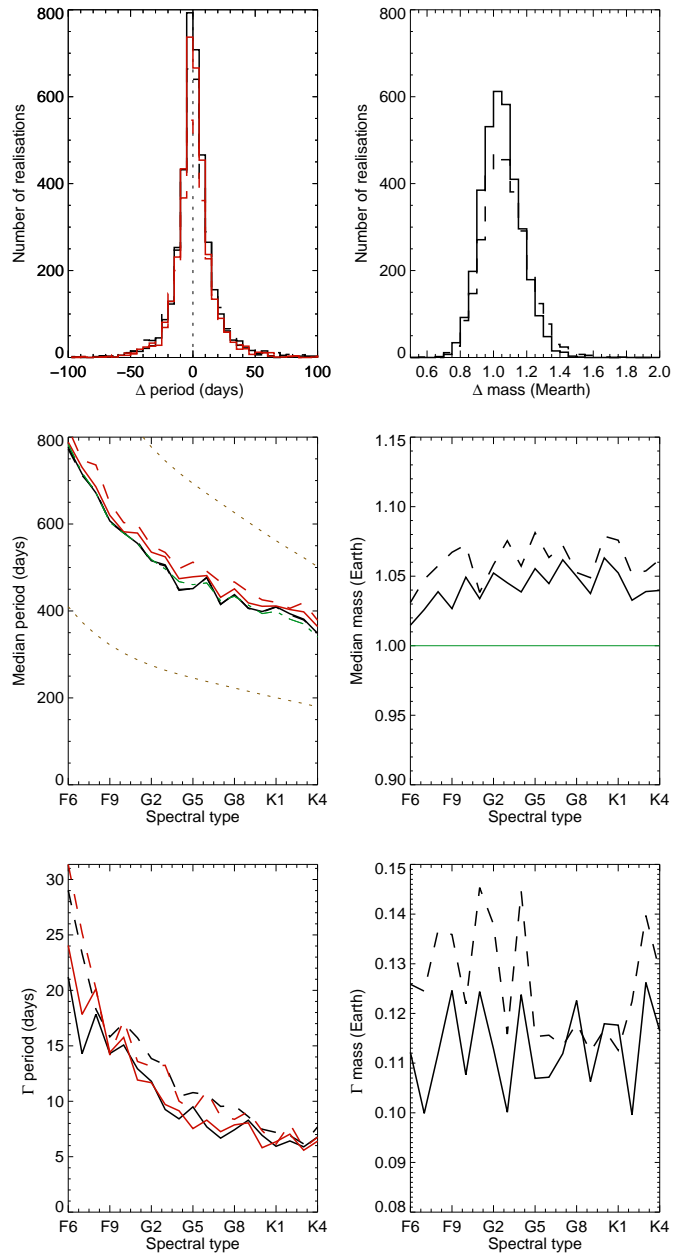}
\caption{
{\it Upper panels}: Distribution of period (left) and mass (right)  difference (between found and true period) for $\Delta$T$_{\rm spot1}$ (solid line) and $\Delta$T$_{\rm spot2}$ (dashed line) from blind tests A. The period plots are for the period corresponding to the highest peak in the periodogram (black) and fitted on the temporal series (red).
{\it Middle panels}: Median period and mass vs. spectral type. The true value is shown in green. The dotted lines in the period plot indicate the boundaries of the habitable zone we consider. {\it Lower panels}: Distribution width (period difference and mass difference) vs. spectral type. 
}
\label{btA_distrib}
\end{figure}

\begin{table*}
\caption{Average detection rates and poor recovery rates (\%)}
\label{tab_taux_moyen}
\begin{center}
\renewcommand{\footnoterule}{}  
\begin{tabular}{lllll}
\hline
Type of detection & $\Delta$T$_{\rm spot1}$ & $\Delta$T$_{\rm spot2}$  & $\Delta$T$_{\rm spot1}$ & $\Delta$T$_{\rm spot2}$ \\ 
  &   FAP 1\% & FAP 1\% & FAP 0.1\% & FAP 0.1\%  \\ \hline
Good recovery (no injected planet) & 99.6 & 100 & 99.6 & 100 \\
Good recovery (injected planet) & 92.5 & 76.7 & 80.0  & 60.1\\
False positive (no injected planet)  & 0.4 & 0 & 0.4 & 0 \\
Incorrect planet & 0.5 &  0.5 & 0.7 & 0.4 \\
Missed  planet & 0.4& 0.5 & 3.2 & 3.4\\
Rejected  planet & 6.6& 22.4 & 16.2& 36.1 \\
\hline
\end{tabular}
\end{center}
\tablefoot{ The rates are averaged over all spectral types. The dependence on spectral type is detailed in Appendix F. }
\end{table*}

\begin{figure}
\includegraphics{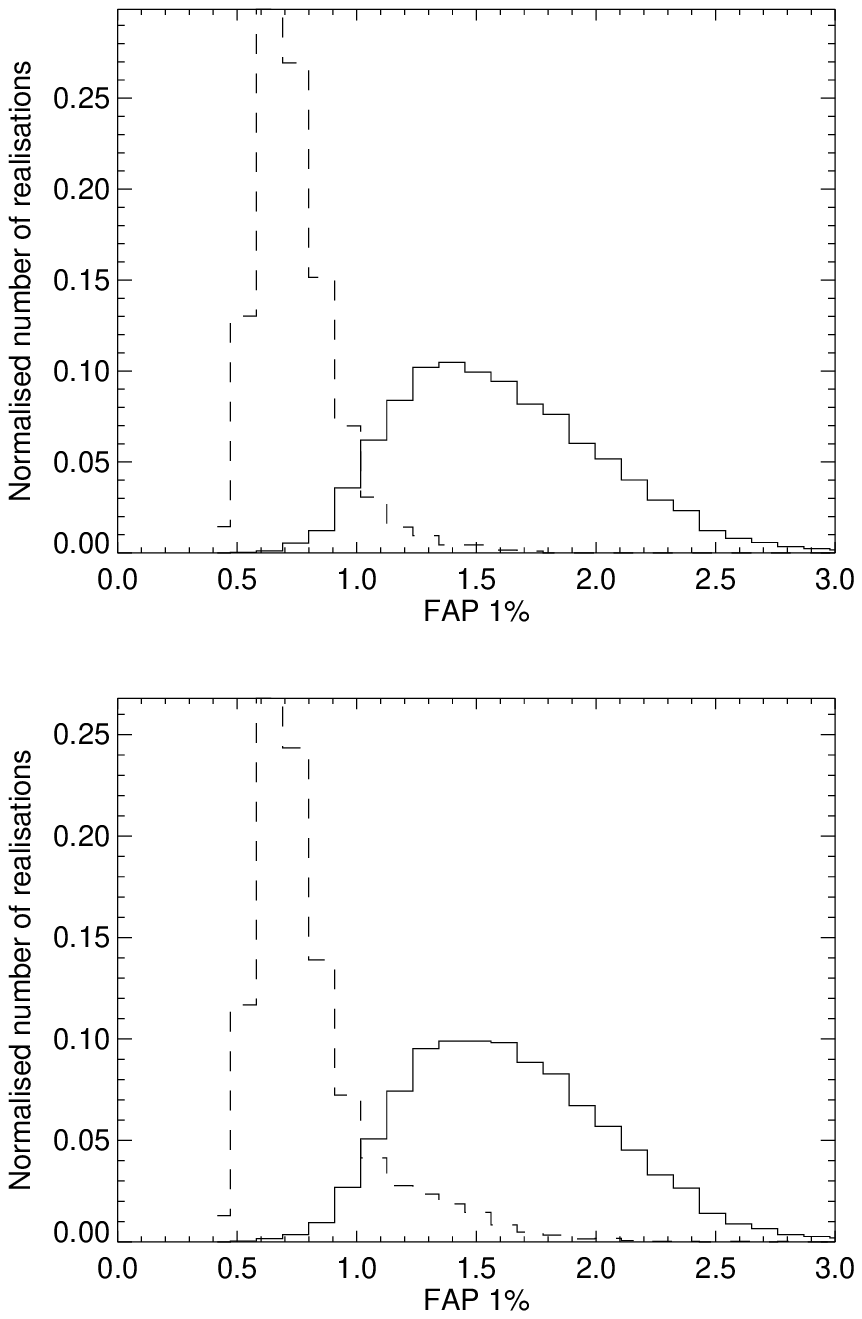}
\caption{
Distribution of FAP values for $\Delta$T$_{\rm spot1}$(upper panel) and $\Delta$T$_{\rm spot2}$ (lower panel) when a planet is injected (solid line) and when no planet is injected (dashed line). 
}
\label{btA_fap}
\end{figure}

\subsubsection{Parameter distributions}

We analysed the difference between the planet parameters obtained in the analysis (period and mass) with the true parameters of the injected planet. The results are summarised in Fig.~\ref{btA_distrib}. The typical width at half-maximum of the period difference distribution is of the order of 20 days. We note that  in a few cases, the periods diverge during the fit, that is, the highest peak in the periodogram is close to the true planet period, but after the fit, it has converged at a completely different peak. This divergence is extremely rare (the two estimate periods differ by more than 100d in less than 0.2\% of the cases), however, and the rms of the difference between the two period estimates is only 10 days. This does not seem to be due to the window function because the fitted period is not closer to the highest peak in the periodogram of that function.

The full width at half maximum for the mass distribution is only of the order of 0.2 M$_{\rm Earth}$ . The dispersion does not vary significantly with spectral type and lies in the range 10-14\%, which is a very good performance. The larger uncertainty is observed for $\Delta$T$_{\rm spot2}$. The bias on the mass of the order of 5-8\% on average is significant, however, as we showed in Sect.~4.1. These properties are close to what we obtained in Sect.~4.1 with the theoretical false positive level of 1\%.

\subsubsection{Detection and false positive rates}

 The average rates are shown in Table~\ref{tab_taux_moyen}.
For an FAP level of 1\% and $\Delta$T$_{\rm spot1}$,  most planets are recovered, while the false positive rate when no planet is injected is very close to zero.   For $\Delta$T$_{\rm spot2}$, the results are still excellent when no planet is injected and  the detection rates are slightly lower when a planet is injected (99.6\% and 80\% on average, respectively). The losses are mostly due to rejected planets (second row of the figure), that is, due to the highest peak at the proper period but below the FAP, and to a lesser extent, the losses are due to missed planets (incorrect period and below the FAP.) We note in particular the very low false positive rates when either no planet is injected (0.4\%  for both $\Delta$T$_{\rm spot1}$ and $\Delta$T$_{\rm spot2}$) and when a planet is injected (0.5\% and 0.7\%): they are below the 1~\% rate expected from the FAP and therefore correspond to a better performance than expected.  There is no significant trend with spectral type (Fig.~\ref{btA_taux}).  
For an FAP level of 0.1\%, the false positives are very rare (0\% when no planet is injected and 0.4-0.5\% when a planet is injected), but the rejected planet rates strongly increase when a planet is injected (22.4\% and 36.1\% for $\Delta$T$_{\rm spot1}$ and $\Delta$T$_{\rm spot2}$ , respectively). 
This FAP level is therefore not very interesting because the effective  false positive level with the FAP at 1\% is already very  low and not improved here. 

Figure~\ref{btA_fap} shows the distribution of the FAP levels when no planet is injected compared to when a planet is injected. There is little difference between the two spot contrasts (the FAP levels are higher for  $\Delta$T$_{\rm spot2}$ by a few percent, as expected). However,  the presence of a planet strongly affects the FAP (the ratio of the median with planet divided by the median without planet is about 2.2), although the objective of the FAP is to provide an estimation of the false positive level due to all processes excluding the planet. This strong overestimation of the FAP when a planet is present probably explains the strong rejected planet rate in Fig.~\ref{btA_taux}. The FAP is therefore a poor estimation of the false positive level for this type of time series, although it is conservative. We note that the sum of the good and rejected planets (i.e. highest peak at the proper period) represents 99.1\% of the realisations for  $\Delta$T$_{\rm spot1}$ and 96.1\% for  $\Delta$T$_{\rm spot2}$, so that if the false positive level could be better estimated, the detection rates would be excellent.

\subsubsection{Relationship between detection and simulation parameters}

\begin{figure}
\includegraphics{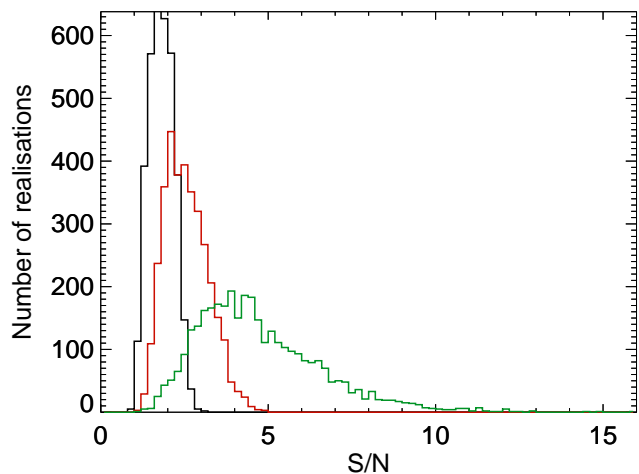}
\caption{
 Distribution of S/N for $\Delta$T$_{\rm spot1}$ and detected planets: global S/N (black), peak S/N (over the whole range, red), and peak S/N (over period higher than 100 days, green). 
}
\label{btA_distribsn}
\end{figure}

\begin{figure}
\includegraphics{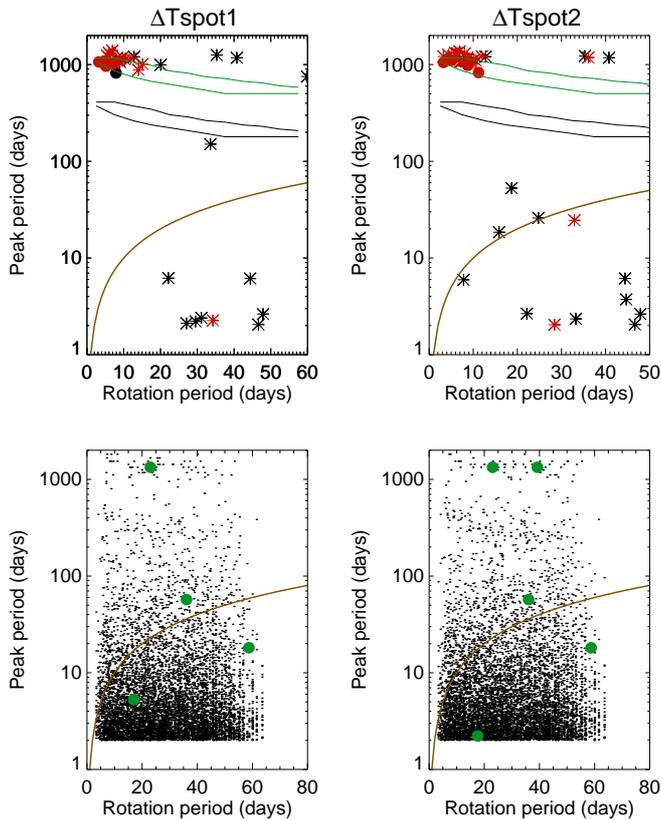}
\caption{
Period of the maximum peak in the periodogram vs. rotation rate for $\Delta$T$_{\rm spot1}$ (left panels) and $\Delta$T$_{\rm spot2}$ (right panels) and in two cases. The first row represents the false positive level when no planet is injected (black) and when a planet is injected (red). Stars correspond to peaks outside the habitable zone and circles to peaks inside the habitable zone. The brown  line indicates the position of the rotation rate. The two black curves represent the extent of the position of the inner side of the habitable zone (for different spectral types), and the green curves show the position of the outer side of the habitable zone. 
The second row represents the second highest peak in the periodogram, those in green indicate the peaks above the FAP. The brown  line is similar to the first row. 
}
\label{btA_prot}
\end{figure}

We detect most planets even with global S/N as low as 1.4-1.6, and  all injected planets are retrieved for S/N above 1.6-1.8, depending on the spot contrast. This means that it may not be necessary to impose a very high S/N such as 5 or 6 to obtain excellent performance (we recall that our false positive rates are very low in these conditions, of the order of 0.5\%, and even lower in the habitable zone, as we show below). The reason is that the global S/N as computed above does not take the frequency behaviour of our signal into account: stellar activity mostly affects short periods (around P$_{\rm rot}$), while the planet is responsible for a peak at much longer periods. We therefore computed  S/N estimates  corresponding to the peak in the periodogram as a more representative indicator: this peak S/N is defined as the ratio between the amplitude of the peak and the amplitude of the highest of the remaining peaks in the whole period range considered (2-2000 days) and for periods longer than 100 days. The distribution of these values is shown is Fig.~\ref{btA_distribsn} for $\Delta$T$_{\rm spot1}$. This peak S/N reaches about 5 for all periods (median of 2.5) and about 16 for periods longer than 100 d (median of 4.4), so that the effective S/N  is much higher than 1. We find that all planets are retrieved for a peak  S/N (period above 100d) higher than 5.

Most false positive either have very short or very long periods (above 1000 days). The latter case is probably due to  the limited length of the time series (3.5 years). There may be a trend for a higher number of false positives for more active stars (evaluated with different criteria), but it is not very significant given their low number.
Their periods are compared to the rotation period in Fig.~\ref{btA_prot}: they are never close to P$_{\rm rot}$ (except for four points in the $\Delta$T$_{\rm spot2}$ case), and otherwise lie at short periods (below 10 days) or close to 1000 days as noted before. An important conclusion  is that 
they do not correspond to planets in the habitable zone, with only nine points there out of our 15200 simulations:  the false positive rate in the habitable zone is therefore extremely low (0.06\%) and  much below the 1\% FAP rate we used to perform the analysis. This confirms that the FAP is strongly overestimated. Furthermore, when no planet are injected, only one false positive out of 15200 simulations is detected, that is 0.007\%, so that false positives are most of the time due to an injected planet at a different period  (even if the planet peak is not significant).

Finally, we also examined the properties of the second highest peak. In most cases, it is below the FAP (except for only nine simulations). Most points lie either at a short period but are not associated with P$_{\rm rot}$ or around 1000 days, as shown in the lower panels of Fig.~\ref{btA_prot}.


\section{Conclusion}

We have used a large set of synthetic time series of complex (solar-like) activity patterns for F6-K4 old main-sequence stars to evaluate their effect on exoplanet detectability in astrometry. We focused on Earth-mass planets orbiting in the habitable zone (and therefore on high-precision astrometry) around such stars and implemented different complementary approaches to assess detectability.  The comparison between the different approaches is provided  in Appendix G.

The rms of the  astrometric  activity time series  in the X direction can be up to two to four times the solar (edge-on) value, depending on spot contrast assumption and inclination, and up to four to five times in the Y direction. The time series exhibit a strong skewness in the Y direction that strongly depends on inclination, which means that the statistical properties of the time series contain information on stellar inclination. However, the presence of noise prevents us from determining this useful stellar parameter. We compared the rms of the astrometric signal to other observables and in particular with photometry and radial velocities. Radial velocities are affected by other processes that we did not include in this comparison, but the knowledge of the stellar variability should allow us to derive a range of astrometric variability (within typically a factor two), and possibly adapt the observational strategy (number of visits for example). 

We found that detection rates for 1 M$_{\rm Earth}$ planets in the habitable zone  are  very good because a large fraction of such planets are expected to be found, with rates above 50\%, which is much better than the rates expected for the RV, especially in the middle of the habitable zone and beyond, for conditions similar to those proposed in the THEIA mission (only 50 visits over 3.5 years) for stars at 10 pc, if technological challenges can be overcome to reach high-precision astrometry. This technique is therefore much more suitable than RV to detect such planets in the range of spectral type and age we considered.  By comparison, the detection rates obtained using a simple method based on the same set of simulations for RV are far lower and very close to 0\% for G stars even with a very large (several thousands) number of observations \cite[][]{meunier19b}. The  presence of granulation and supergranulation in RV also leads to poor detection rates based on  blind tests similar to those performed in the present paper, usually below 50\% \cite[][]{meunier20b}.  We expect RV performance to be slightly better for K stars than for F-G stars \cite[e.g.][]{meunier19b}, so that astrometry could be very interesting for the latter category of stars, where many stellar processes significantly affect the RVs.  Furthermore, the uncertainty on the fitted periods, and more importantly, on the mass are very good. The uncertainties on the mass are indeed below 20\%, which would be very interesting for PLATO follow-ups. 

The application of the classical bootstrap method to estimating the FAP in our blind tests  led to important conclusions. First, the FAP is significantly overestimated, especially when a planet is present (by more than a factor 2), so that the false positive level is very low, especially in the habitable zone. However, for the same reason, some planets remain undetected even though the highest peak most of the time is at the true planetary period. Better indicators of the false positive should then be used, for example, using our theoretical knowledge of the stars as shown by the theoretical false positive levels obtained in this paper. A possible strategy to obtain better results would also be to make more visits to certain stars: doubling the number of points provides rates close to 100\%.
 We also observed that it is possible to obtain very good rates with a low level of false positives ($<$0.5\%) for a global S/N between only 1 and 2 because this indicator  does not represent the S/N of the peaks in the periodogram well, for which the S/N is much higher.

In a future work, we will use this large set of synthetic time series and the approaches described in this paper to re-evaluate the expected detection rates and detection limits of the stars in our neighbourhood. They constitute the main targets of future high-precision astrometry missions.

\begin{acknowledgements}

This work has been funded by the ANR GIPSE ANR-14-CE33-0018.
We are very grateful to Charlotte Norris who has provided us the plage contrasts used in this work prior the publication of her thesis. 
We thank Pascal Rubini for his work to convert the simulation code in C++. 
This work was supported by the "Programme National de Physique Stellaire" (PNPS) of CNRS/INSU co-funded by CEA and CNES.
This work was supported by the Programme National de Plan\'etologie (PNP) of CNRS/INSU, co-funded by CNES.

\end{acknowledgements}

\bibliographystyle{aa}
\bibliography{38710corr}

\begin{appendix}

\section{Planetary orbits}

The planetary orbit for the first category of computations (planet seen edge-on, aligned with the X-axis of the star) is described as \cite[][]{eisner01}

\begin{equation}
 x_{\rm pla}=A_c \cos(\omega t) + A_s \sin (\omega t)  
,\end{equation}

where $A_c$=$\alpha \sin \phi$ et $A_s$ = $\alpha \cos \phi$ ($\phi$ related to the phase of the planet on its orbit), and $\omega$ is 2$\pi$/P$_{\rm pla}$. \\

\begin{figure}
\includegraphics{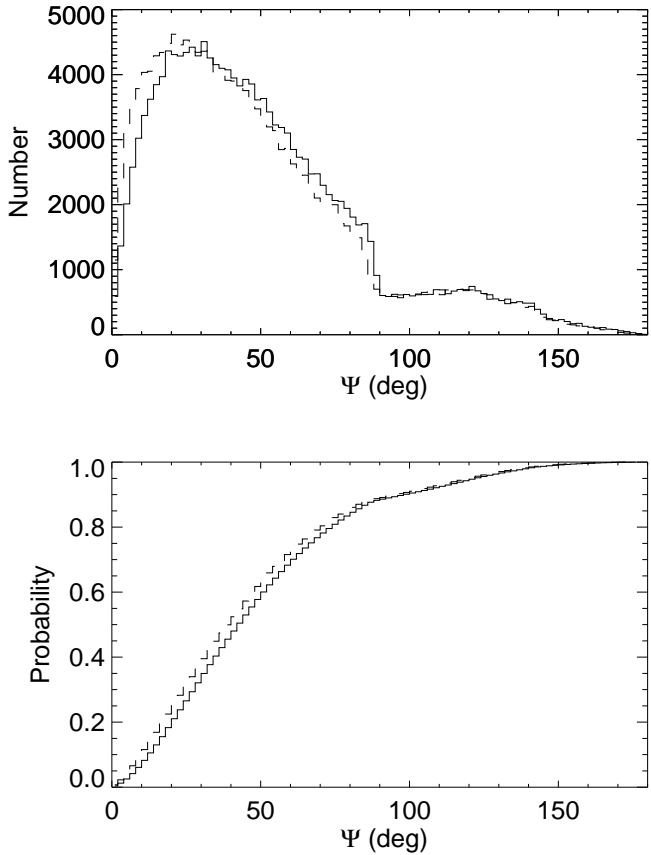}
\caption{{\it Upper panel:} Distribution of the angle between the stellar equatorial plane and the planet orbital plane $\Psi$ assuming that transiting planets are seen edge-on seen  (solid line) and assuming that transiting planets are seen  with a small dispersion around the edge-on configuration (dashed line). 
{\it Lower panel:} Same for the cumulated distribution. 
}
\label{psi}
\end{figure}

For the more complete approach, we considered two assumptions  because the angle $\Psi$ between the planetary orbital plane and the stellar equatorial plane is poorly constrained: 
\begin{itemize}
\item{ 
{\it A. Distribution of the angle $\Psi$.}  These choices result from the observation that there are known exoplanets with  values of $\Psi$ significantly different from 0: we expect $\Psi$ to affect the relationship between the astrometric signal in the X and Y directions.  We used the results obtained using the Rossiter-McLaughlin effects (obtained for more massive transiting planets). Very few studies provide the true angle, and in general, only the projected (on the sky) angle is provided. We therefore used the list of projected obliquities provided in the TEPCat catalogue \cite{southworth11}\footnote{Orbital obliquity observations for transiting planetary systems, http://www.astro.keele.ac.uk/jkt/tepcat/obliquity.html}, and then performed a Monte Carlo simulation to derive a realistic distribution of true obliquities corresponding to these projected obliquities, following  the procedure described in  \cite{fabrycky09}: we assumed a planetary orbit seen edge-on (because they are seen to be transiting in this catalogue\footnote{Assuming a distribution close to the edge-on configuration does not significantly change  the results.}) and a uniform distribution in cos $i$. The resulting distribution in $\Psi$ is shown in Fig.~\ref{psi}.  Such an approach has been used in previous works \cite[e.g.][]{triaud10,brothwell14}.  }
\item{ {\it B. $\Psi$=0 (i.e. the orbital plane and the equatorial plane are the same)}. This simple assumption has been used in several works \cite[e.g.][]{meunier20b}, and low values of $\Psi$ correspond to telluric planets in the Solar System. }
\end{itemize}

The planetary orbits were then modelled as follows. The orbit was first described in the orbital plane of the planet as \cite[][]{eisner01}

\begin{equation}
x_{\rm pla}=A_c \cos(\omega t)+A_s \sin (\omega t) 
\end{equation}
\begin{equation}
 y_{\rm pla}=-A_s \cos(\omega t)+A_c \sin (\omega t)  
.\end{equation}

We then perform several successive projections: following the true obliquity $\Psi$ (either 0 or following the distribution described previously), then following the angle between the line of sight  and the node line $\Phi_0$ (chosen randomly), and then following the stellar inclination $i$. This leads to the coordinates of the planet in the plane of the sky, 

\begin{equation}
 x_{\rm pla}'=A' \cos(\omega t)+B' \sin (\omega t)  
\end{equation}
\begin{equation}
 y_{\rm pla}'=A'' \cos(\omega t)+B'' \sin (\omega t) 
,\end{equation}
where $x_{\rm pla}'$ corresponds to the X direction of the star, and where 
\begin{equation}
 A'= -A_s \cos(\Psi) \cos \Phi_0 + A_c \sin \Phi_0 
\end{equation}
\begin{equation} B'= A_c \cos(\Psi) \cos \Phi_0 + A_s \sin \Phi_0 
\end{equation}
\begin{equation}
 A'' = -A_s \sin(\Psi) \sin (i) +(A_s \cos(\Psi) \sin \Phi_0 + A_c \cos \Phi_0 ) \cos (i)  
\end{equation}
\begin{equation}
 B'' = A_c \sin(\Psi) \sin (i) -(A_c \cos(\Psi) \sin \Phi_0 -A_s \cos \Phi_0)   \cos (i) 
.\end{equation}

\section{False positive levels}

We first detail   the criteria we used to compare the signal with an injected planet with a false positive level, and then  the compute the false fp levels from the simulations for each spectral type. Then we show effect of the dependence of the fp level on activity and inclination on the results.

\begin{figure}
\includegraphics{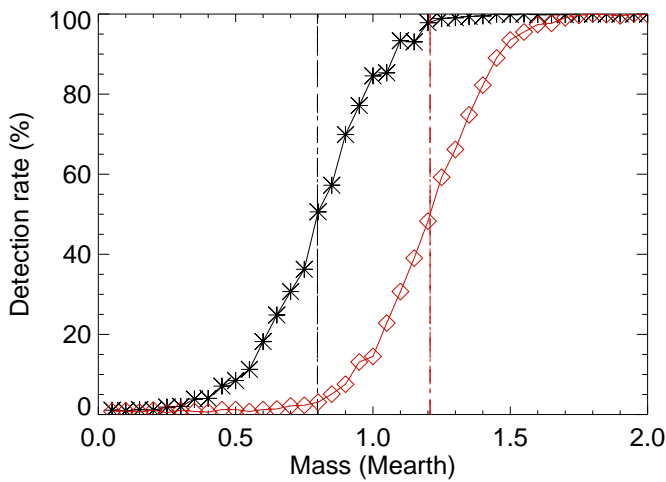}
\caption{
Detection rate (above 1\% false positive level) vs. mass for G2 stars,  PHZ$_{\rm in}$,  and $\Delta$T$_{\rm spot2}$ for the frequential (black) and temporal analysis (red). The vertical lines indicate the position of the 50\%  detection rate levels.}
\label{taux}
\end{figure}

\subsection{Criteria}

Two types of criteria were used in this paper.  
First, we performed  a frequential analysis \cite[][]{marcy05,catanzarite08,makarov09,traub10,makarov10}: we computed the maximum power in the periodogram at periods close to the orbital period we wished to characterise, first without a planet to compute the false positive level, and then with a planet to compute the detection rate by comparison with the false positive level computed over a large sample of simulations. Because we are interested in the false positive level in a  certain domain (habitable zone), we considered (PHZ$_{\rm in}$, PHZ$_{\rm med}$, and PHZ$_{\rm out}$) for each of the orbital periods, the maximum of the periodogram was computed in the range [0.5P,2P], where P takes one of these three values (the results are not very sensitive to the exact range).   

Second, a direct fit of the signal, using a $\chi^2$ minimisation, can be made, assuming a planetary signal, as in \cite{eisner01,eisner02}, who claimed that this should constitute a more robust approach. The model used to perform the fit is either the model described in equation 1 (for the simple edge-on configuration) or using equations 4-5 (and then using equations 6-9 to retrieve the original parameters, and then $\alpha$ and the planet mass). An offset was also added to the model. We note that the main objective here was not to refine the minimisation technique but to analyse a very large number of synthetic time series using standard tools. Unless indicated otherwise, the stellar inclination $i$ was assumed to be known: this should not affect the fitted amplitude and period significantly\footnote{A dedicated blind test similar to those performed in Sect.~4 shows that the results are not significantly impacted.} because its effect should be on the angles $\Psi$ and $\Phi0$, and there is a strong degeneracy between these three angles in any case \cite[see discussion in][]{eisner01}. We  estimate the false positive level separately for each spectral type and spot contrast. This corresponds to 6480 simulations in most cases (except for K stars for which the number of simulations is slightly lower). 

We used a similar approach using the RV technique to estimate the effect of granulation and supergranulation on exoplanet detectability \cite[][]{meunier20b}. The situation is more complex here, however. In \cite{meunier20b}, we considered a number of realisations, for example, of granulation, for a given spectral type, but  all simulations corresponded to the same amplitude of the signal. In the present case,  simulations of a given spectral type correspond to different activity levels, both on average and temporal variability. They also have different inclinations, which could correspond to different stellar signal amplitudes.   The details of the computation are presented in the following section, as well as the effect of a variable fp determination corresponding to a given spectral type on the results. In most of the paper, we used a constant fp level for a given spectral type. We show that the effect is minor because the detection rates are very close to 100\% in many configurations. 

 Finally,   using a similar approach but considering different masses, it is possible to compute the mass (i.e. the detection limit) corresponding to a given detection rate (e.g. 50\% or 95\%), for the same 1\% false positive rate. This is illustrated in Fig.~\ref{taux}. We note that the frequential approach leads to lower detection rates than the temporal approach (for a given mass).    The reason is not yet clear, but it may be due to a difference in sensitivity to the noise because we see, for example, in Sect.~4.1.1 that the noise leads to a bias in the mass estimation so that the temporal method is sensitive to it.

\subsection{Computation of false positive levels}

\begin{figure}
\includegraphics{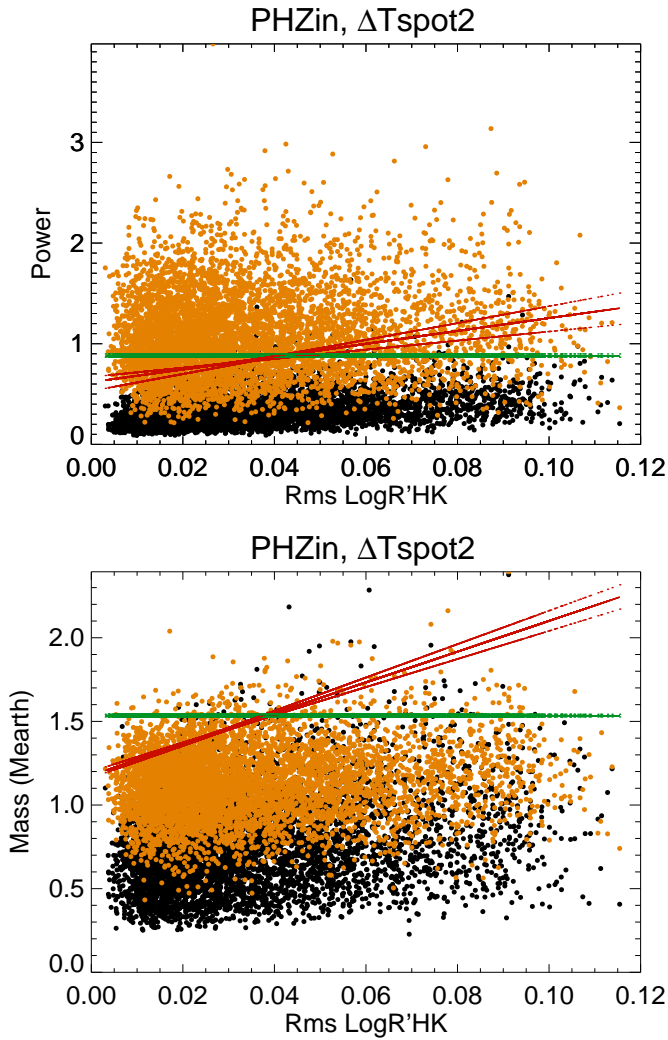}
\caption{
Illustration of fp computation for K2 stars, all inclinations, $\Delta$T$_{\rm spot2}$,PHZ$_{\ in}$, representing power (frequential analysis, upper panel) and fitted mass (temporal analysis, lower panel) vs. rms of $\log R'_{HK}$ for no planet (black) and with an injected planet (orange, 1 M$_{\rm Earth}$). The green line corresponds to a constant fp level of 1\%, and the red line shows a fp level depending on the rms of $\log R'_{HK}$. The green and red dots show the upper and lower bounds corresponding to the criterion (see text).  }
\label{ex_fp}
\end{figure}

\begin{figure}
\includegraphics{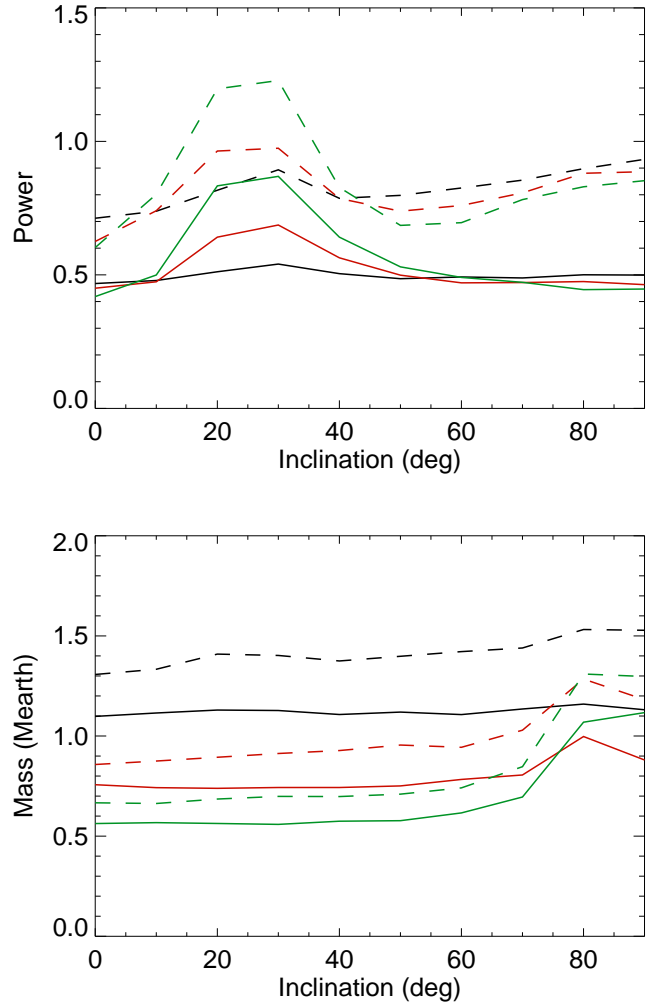}
\caption{
False positive fp  averaged over all simulations vs. inclination for different PHZ (inner side in black, middle in red, and outer side in green) and for $\Delta$T$_{\rm spot1}$  (solid line) and $\Delta$T$_{\rm spot2}$ (dashed line) for the frequential analysis (upper panel) and temporal analysis (lower panel).}
\label{fp_incl}
\end{figure}

 In these simulations, the stellar activity level should affect the fp level:

\begin{itemize}
\item{If we consider a single value of the fp level for a given spectral type: this means that we consider all stars of this spectral type as equivalent because we do not have any information on the activity level of the star, for example. We are then interested in the probability that a planet can be detected around a star of that spectral type globally, that is, knowing only  its spectral type.  }
\item{However, we expect that if a star is quiet, then it should be easier to detect a given planet around it, but if the star is active, it should be more difficult, that is, the fp level might in fact be slightly different. If the variability of the star is well characterised, then we can  estimate the fp level by separating stars that are strongly variable from those that are not.  }
\end{itemize}
The same argument might be made by separating simulations according to their inclination. 

Figure~\ref{ex_fp} illustrates the computation of the fp level for a K2 star, $\Delta$T$_{\rm spot2}$, and the inner side of the habitable zone using the frequential approach (upper panel) and the temporal approach (lower panel). The black dots correspond to the amplitude (either the power or the mass) when no planet is injected, which we used to compute the fp level: the green horizontal line is the fp level of 1\% when it is considered constant for that spectral type, that is, 1\% of the black dots are above the green line. The orange dots, corresponding to the same simulations with a 1 M$_{\rm Earth}$ mass planet injected, can then be used to compute the detection rate corresponding to this 1\%  fp level. 

The red line, on the other hand, provides the fp level when a dependence on the rms of $\log R'_{HK}$, hereafter $r$,  is considered, that is, based on the amplitude of the stellar variability. It is computed as follows: we defined  bins in $r$ and assumed that fp follows a linear trend in $r$. We computed 1000 simulations of the slope, adjusting the straight line so that the percentage of black dots above the line was as close as possible to 1\% in each bin. We kept the best solutions (i.e. that met the condition for the maximum number of bins, usually 4). The median of those solutions is shown as the red line in Fig.~\ref{ex_fp}. With this computation, the detection rates should be better for the less variable stars than  for the most variable stars. However, we note that this is observed mostly for the configuration shown in this figure ($\Delta$T$_{\rm spot2}$, inner side of the habitable zone): for the other configurations, there is not much difference between the two
choices (constant or variable fp level) because the detection rates are much better and in fact at the 100\% level (most orange dots are well above the black dots). The effects of this choice is investigated in   Appendix B.3. 

We also studied how  the false positive levels depend on inclination. For each orbital period (three values in the habitable zone) and $\Delta$T$_{\rm spot}$, we averaged the  fp levels over all simulations of a given spectral type and inclination. The results are shown in Fig.~\ref{fp_incl}. A few trends can be seen. There is a bump around 20-30$^\circ$ for the frequential approach (upper panel), mostly for the middle and outer habitable zone, and then a small trend for $\Delta$T$_{\rm spot2}$. This is explained in Sect.~3.1. The curves are mostly flat for the temporal approach (lower panel), except in a few cases that are close to edge-on.

\subsection{Effect of the assumptions made during the false positive level computation}

\begin{figure}
\includegraphics{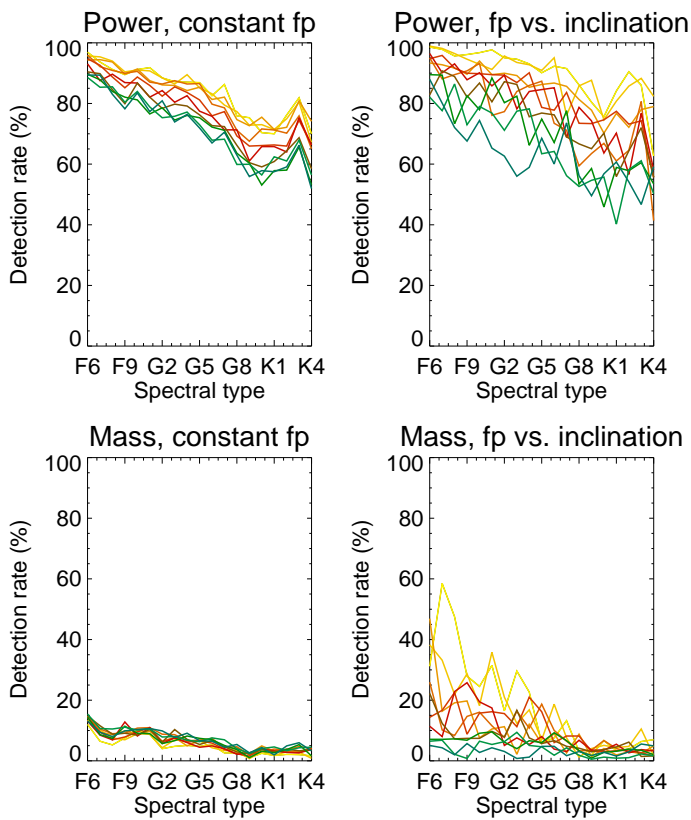}
\caption{
Detection rates vs. spectral type for 1 M$_{\rm Earth}$ planet based for the frequential analysis (upper panel) and temporal analysis (lower panels), separately for  ten inclinations (colour code as in Fig.~3). The detection rates are shown for $\Delta$T$_{\rm spot2}$ and the inner side of the habitable zone. The fp levels depend on inclination for the right panels, but they do not for the left panels.  }\label{incl1}
\end{figure}

\begin{figure}
\includegraphics{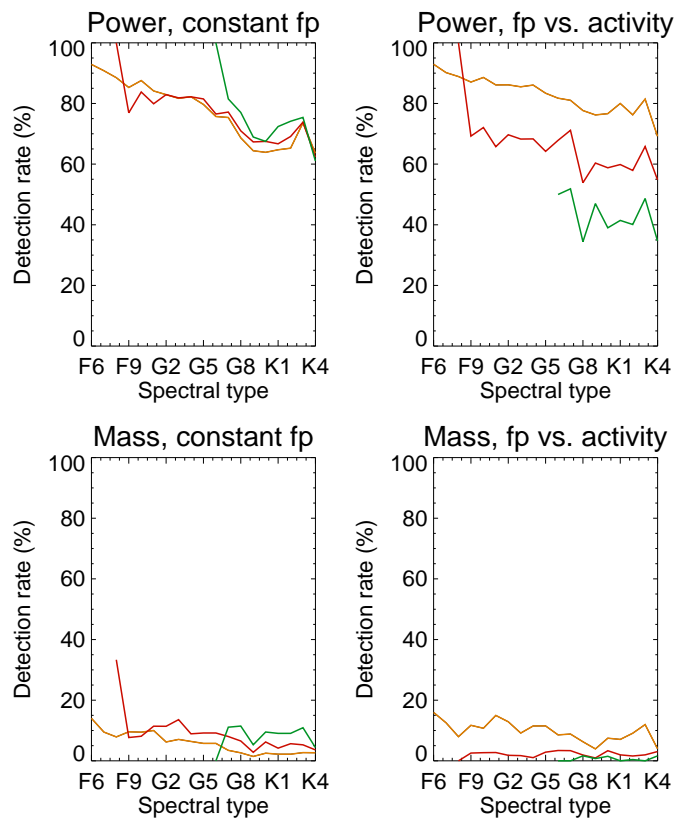}
\caption{
Same as Fig.~\ref{incl1} for fp levels depending on activity variability. The colour code corresponds to different activity variability levels (based on the rms of $\log R'_{HK}$): quiet (orange), intermediate (red), and active (green). }
\label{act1}
\end{figure}

We showed in Sect.~4.2.2 the detection rates corresponding to the theoretical fp level determined for each spectral type. Here, we consider the effect of the activity level and of inclination on this determination and on the subsequent  detection rates.

When a variable level of fp  is used within a spectral type bin (i.e. versus inclination or activity variability), these detection rates are very similar. However, when we consider stars of various activity levels or different inclinations separately, for example, we expect different detection rates. Because they are often very close to 100\%, however, the effect is in fact very small. We illustrate the effect of these variable false positives levels in the case of the inner side of the habitable zone and the high spot contrast, for which the effect is the largest because the detection rates are below 100\%.  
Figure~\ref{incl1} shows the effect of inclination in this configuration. The left panels correspond to the constant fp  level, and on the right side, both fp   levels and detection rates are computed separately for each inclination. When a constant fp  level (i.e. averaged over all inclinations) is considered, the detection rates are higher when the star is seen pole-on. This is expected  because more information is available in this case (from the X and Y directions). The difference between edge-on and pole-on configurations is small, however. When the inclination-dependent false positive level (obtained in Sect.~2.4.2) is considered, the trend is slightly reinforced, so that pole-on configurations are clearly easier targets. 

Figure~\ref{act1} shows a similar analysis  for the dependence on the activity variability. When the same false positives for all simulations within a spectral type bin (left panels) are considered, there is a difference in detection rate depending on the activity level, but it is very small and not significant. When the false positive rates are computed separately for the different simulation categories (criterion based on the rms in $\log R'_{HK}$), however, the detection rates are better for the quiet stars than for the most active stars, as expected (e.g. between  80\% and 40\% for K stars). This gives an estimate of the range in detection rates that is expected for stars of different variability levels. We recall, however, that this is seen mostly for the inner side of the habitable zone and high spot contrasts: for all other configurations, the rates are always very close to 100\% and show little departure from it even for the most active stars.

\section{Properties of the  stellar astrometric signal}

 In this appendix, we show more detailed results concerning the characterisation of the astrometric signal. We first show the effect of inclination on the properties of the astrometric signal,  and discuss the properties for spots and plages separately. The relationship between the properties in the X and Y directions is studied in more detail. Finally, we show in more detail the relationship between the astrometric and the photometric signals. 

\begin{figure}
\includegraphics{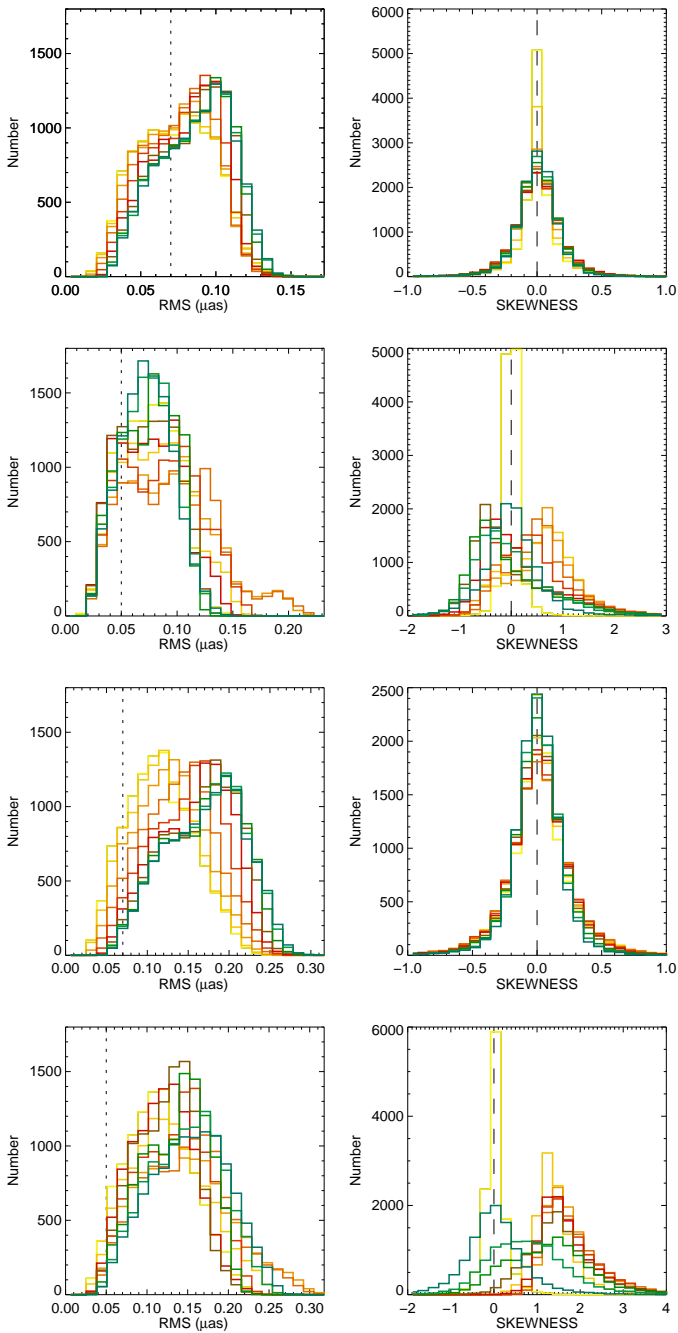}
\caption{
Rms (left) and skewness (right) distributions of the activity time series in both  directions for (from upper to lower panel) the X direction and $\Delta$T$_{\rm spot1}$, the Y direction and $\Delta$T$_{\rm spot1}$, the X direction and $\Delta$T$_{\rm spot2}$, and the Y direction and $\Delta$T$_{\rm spot2}$. Different curves show the ten inclinations (same colour code as in Fig.~1). 
}
\label{act_b}
\end{figure}

\begin{figure}
\includegraphics{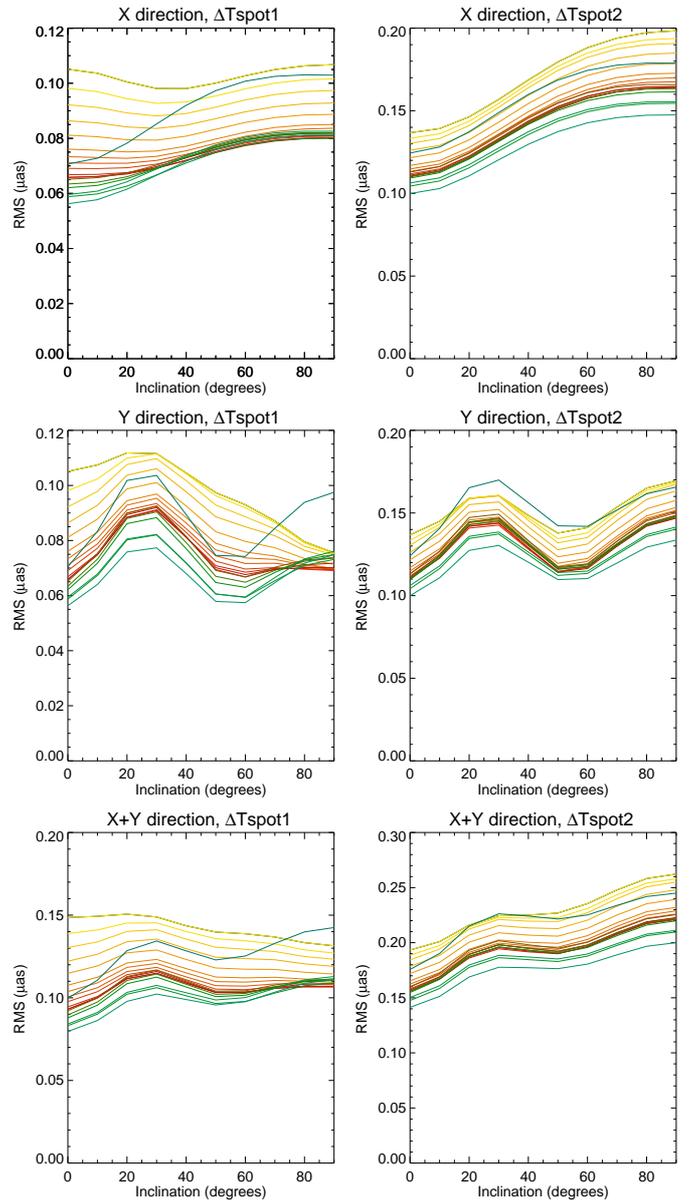}
\caption{
Rms of time series in the X direction (upper panels) and the Y direction (middle panels), and combining both directions (low panels) for $\Delta$T$_{\rm spot1}$ (left) and $\Delta$T$_{\rm spot2}$ (right). The colour code represents the spectral type from F6 (yellow) to K4 (blue). 
}
\label{act_rms}
\end{figure}

\begin{figure*}
\includegraphics{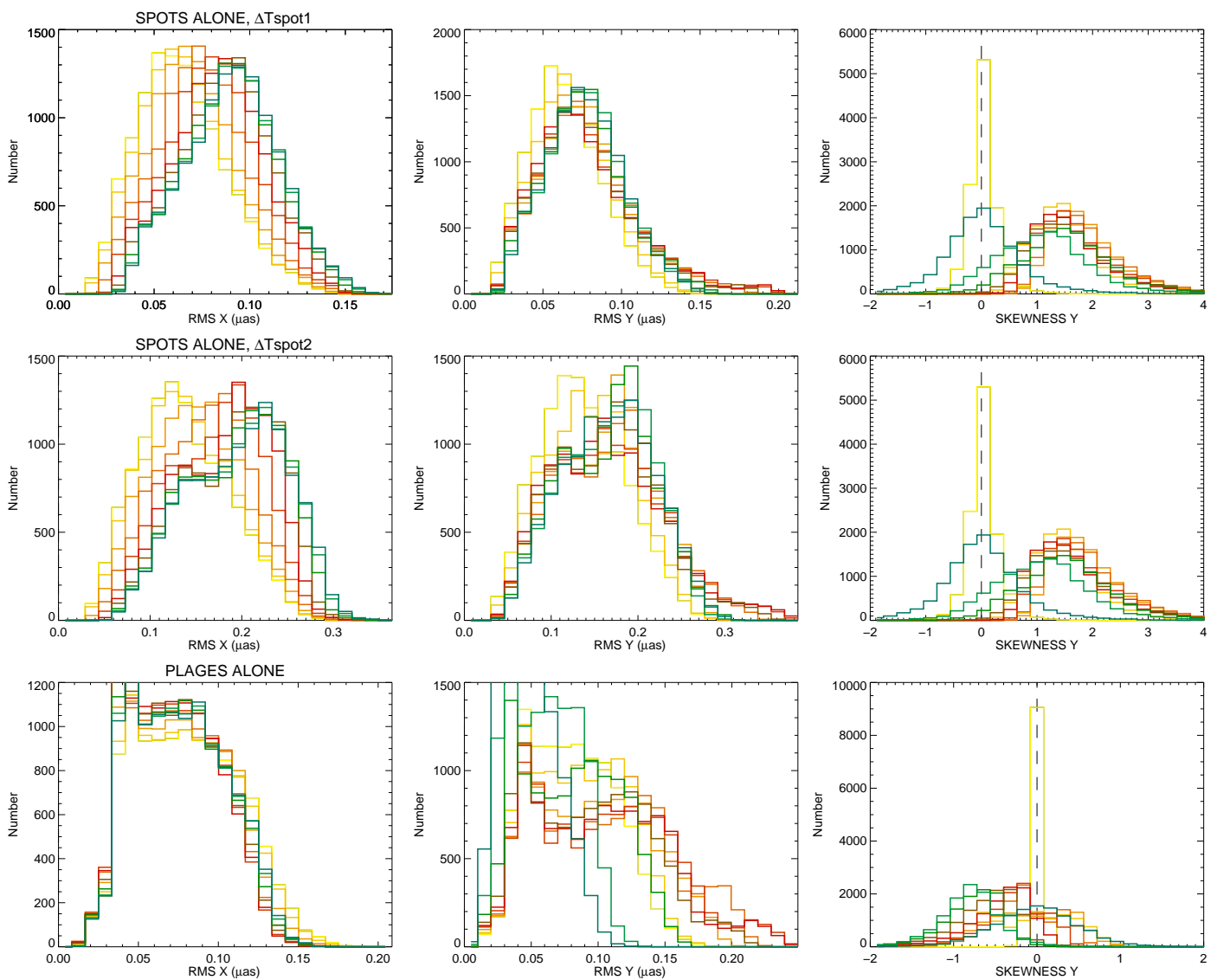}
\caption{
Distribution of rms in the X direction (left), of rms in the Y direction (middle) and of skewness in the Y direction (right) for ten inclinations (the colour code is similar to Fig.~\ref{act}) for different contributions: spots with $\Delta$T$_{\rm spot1}$ (upper panels), spots alone $\Delta$T$_{\rm spot2}$ (middle panels), and plages alone (lower panels). 
}
\label{compo}
\end{figure*}

\subsection{Effect of inclination}

 Figure~\ref{act_b}  shows the distributions of rms and skewness computed for each time series for the different inclinations and all spectral types and activity levels we considered. We observe a trend with inclination, as well as a positive skewness for the Y direction when it is far from pole-on.
Figure~\ref{act_rms} shows the average rms versus inclination separately for the different spectral types. The rms in the X direction is decreasing for increasing inclination (from pole-on to edge-on).  In the Y direction and when the  two directions are combined, there is a bump around 20$^\circ$, followed by an increase in $\Delta$T$_{\rm spot2}$, which may explain how fp depends on inclination  (Sect.~2.4.2).

\subsection{Contribution of spots and plages}

We illustrate separately the contribution of spots and plages to the astrometric signal. The distributions of the rms in the X and Y direction and of the skewness in the Y direction are shown for  spots alone ($\Delta$T$_{\rm spot1}$ in the upper panels and $\Delta$T$_{\rm spot2}$ in the middle panels) and plages alone (lower panels) in Fig.~\ref{compo}. For spots, the inclination effect is more clearly seen for the rms in the X direction and the skewness. For plages, the effect is more visible for the rms in the Y direction and the skewness. The sign of the skewness is  reversed compared to spots. The shapes of the rms distributions are different as well. The rms due to plages is of the same order of magnitude as that for spots alone ($\Delta$T$_{\rm spot1}$ assumption).

\begin{figure}
\includegraphics{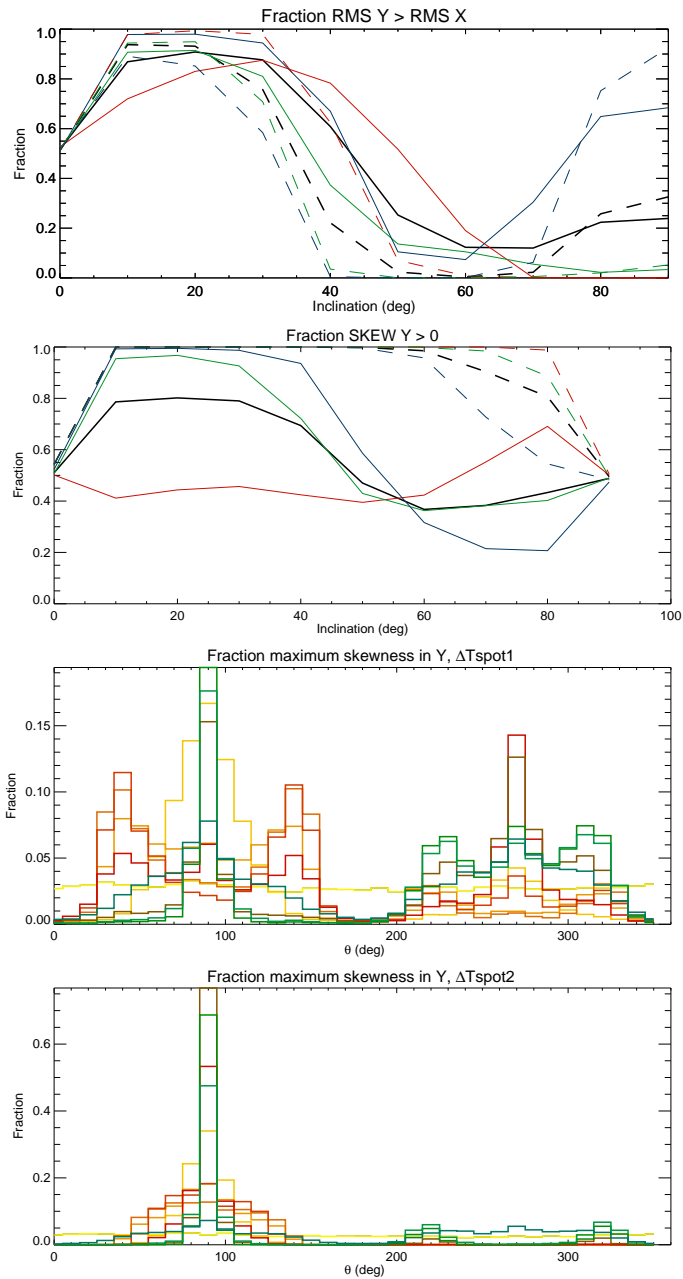}
\caption{
{\it First panel:} Fraction of simulations in which the rms is higher in Y than in X direction for $\Delta$T$_{\rm spot1}$ (solid) and $\Delta$T$_{\rm spot2}$ (dashed). The colour code represents different $\theta_{\rm max}$ values: solar (red), solar+10$^\circ$ (green), solar+20$^\circ$ (blue), and all (black).
{\it Second panel:} Same for the fraction of simulations with a positive skewness in the Y direction.
{\it Third panel:} Distribution of the angles at which the skewness in Y is highest ($\Delta$T$_{\rm spot1}$  only) for the ten inclinations (same colour code as Fig.~3).
{\it Fourth panel:} Same for $\Delta$T$_{\rm spot2}$.
}
\label{fraction}
\end{figure}

\begin{figure}
\includegraphics{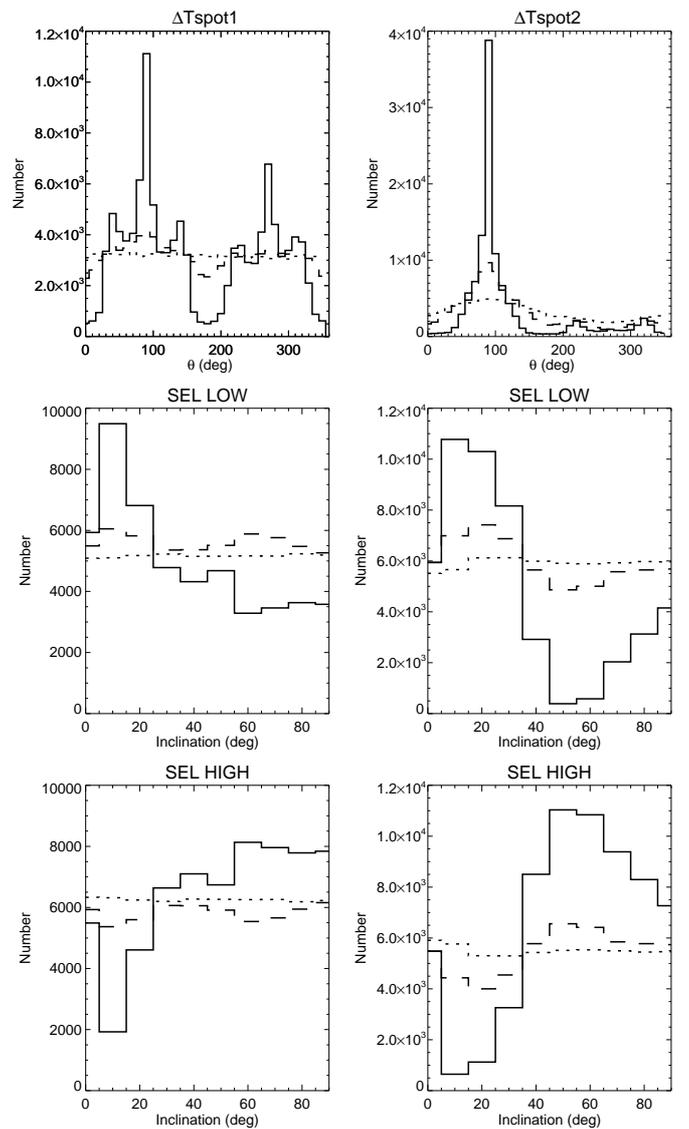}
\caption{{\it Left column, from top to bottom:} Distribution of different variables characterising activity for $\Delta$T$_{\rm spot1}$ and for different time series: complete without noise (solid lines), degraded sampling without noise (dashed lines), and degraded sampling with noise (dotted lines). Variables are the angle at which the skewness is highest (90$^\circ$ and 270$^\circ$ represent the Y direction), inclination for simulations in which the rms in this direction is higher than in the orthogonal direction, and inclination for simulations in which the rms in this direction is lower than in the orthogonal direction.  
{\it Right column:} Same for $\Delta$T$_{\rm spot2}$.
}
\label{protocole}
\end{figure}

\subsection{Relation between the stellar astrometric signal in the X and Y directions}

As already shown in Sect.~3.1 (Fig.~\ref{act}), the stellar astrometric signal in the X and Y directions does not show the same behaviour: the amplitude can be different, with a ratio that depends on inclination,  and the skewness properties are also different. 
Figure~\ref{fraction} illustrates these properties in more detail. All spectral types and activity levels are considered. The first panel shows the fraction of simulations where the rms in the Y direction is higher than in the X direction versus stellar inclination: This fraction is larger for low inclinations (above 50\%) and smaller (below 50\%) for high inclinations. It is sensitive to the latitudinal extent of the activity pattern. The second panel shows the fraction of simulations in which the skewness in the Y direction (which showed an asymmetry in Fig.~\ref{act}) is positive: it is overall larger than 50\%, but only at low inclinations (below 40$^\circ$) for $\Delta$T$_{\rm spot1}$, and it is present for most inclinations for $\Delta$T$_{\rm spot2}$. The two other panels illustrate this from another point of view. The frame of reference is rotated, so that a $\theta$ angle of 0$^\circ$ corresponds to the time series projected on the original X direction, and 90$^\circ$ corresponds to the time series projected on the original Y direction. For each time series projected using various values of $\theta$, we identify the $\theta$ value for which the skewness is highest. A peak in the distribution at 90$^\circ$ , for example, means that the skewness is maximal in the Y direction.  We find that there are indeed peaks at 90$^\circ$ and 270$^\circ$, especially for $\Delta$T$_{\rm spot2}$. For $\Delta$T$_{\rm spot1}$, these peaks are present, but for certain inclinations (intermediate inclinations, curves in red), the skewness is highest for different projections (intermediate between X and Y).

We examined whether such properties might be used to determine the stellar inclination as well as the X and Y directions from a given observation. We first considered the full time series without noise. The protocol was the following. The distribution of $\theta$ values at which the skewness in Y was highest was plotted. We also separated the simulations into two groups, those in which the rms in Y is higher than the rms in X (called SEL LOW),  and those in which the rms in Y is lower (called SEL HIGH). We then compared the inclination distribution for these two categories: if the distributions are well separated, then we expect to be able to use the criterion to identify stars with low or high inclinations. The results are shown as solid lines in Fig.~\ref{protocole}. The inclination distributions for the two selections peak in different domains, although there is an overlap.  When the sampling is degraded, however (dashed lines), the distributions completely overlap, and even more when noise is added (dotted ones). We conclude that it is therefore unlikely that such properties could be used to characterise the geometry of the system. 

\subsection{Relationship between astrometry and photometry}

\begin{figure*}
\includegraphics{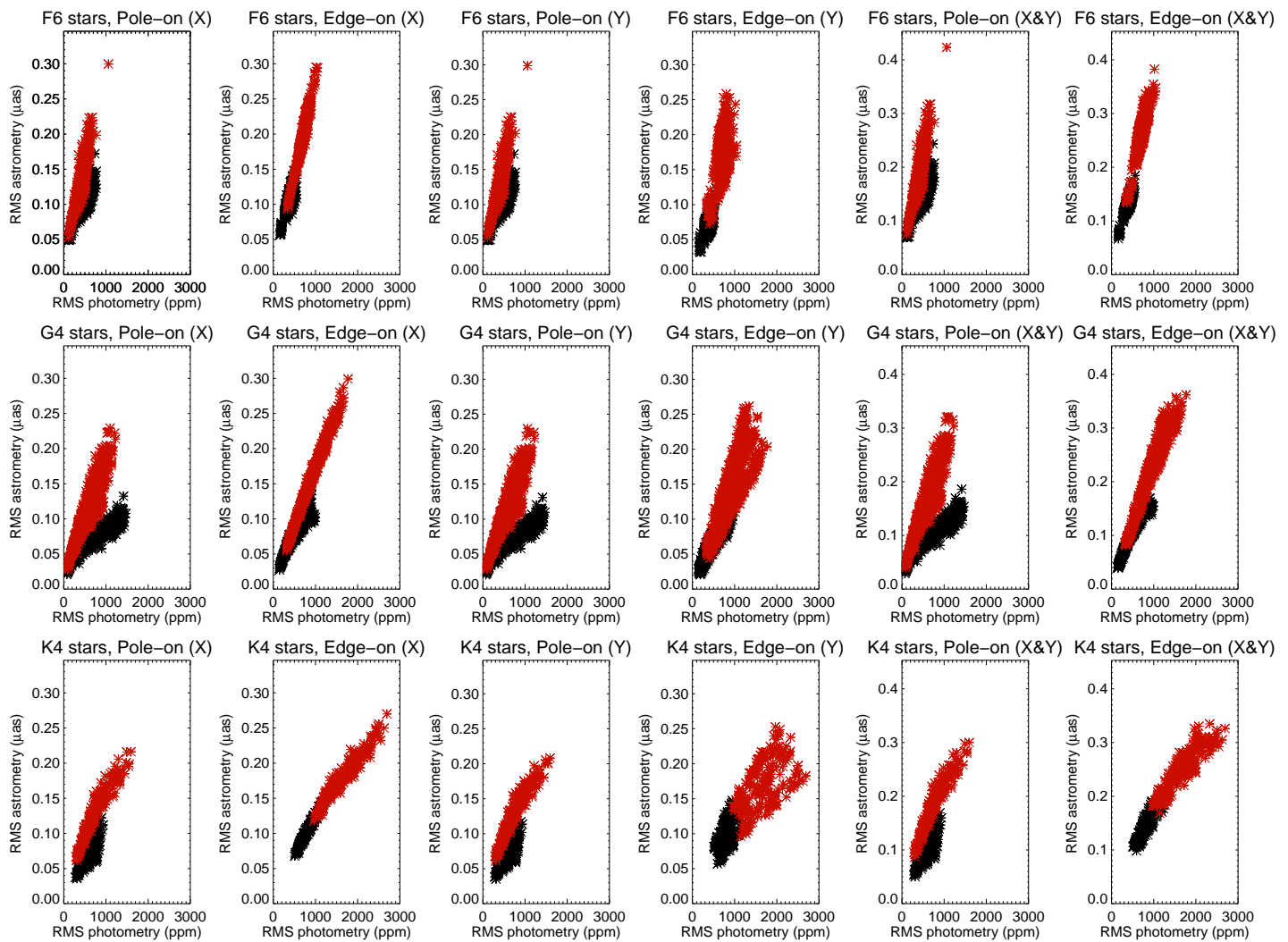}
\caption{
Rms of astrometric signal vs. rms photometry for various configurations (spectral type and inclination) for $\Delta$T$_{\rm spot1}$ (black) and $\Delta$T$_{\rm spot2}$ (red).  }
\label{photom}
\end{figure*}

 Figure~\ref{photom} separately illustrates in more detail the relationship between astrometry and photometry for three different spectral types and two inclinations. This illustrates the typical expected amplitude of the astrometric signal when the photometric variability is known. The relations are not strictly linear because there is saturation at high activity level.  The inclination effect in some of these relations is strong, mostly for the astrometric signal in the Y direction.

\section{Detectability based on S/N}

\begin{figure*}
\includegraphics{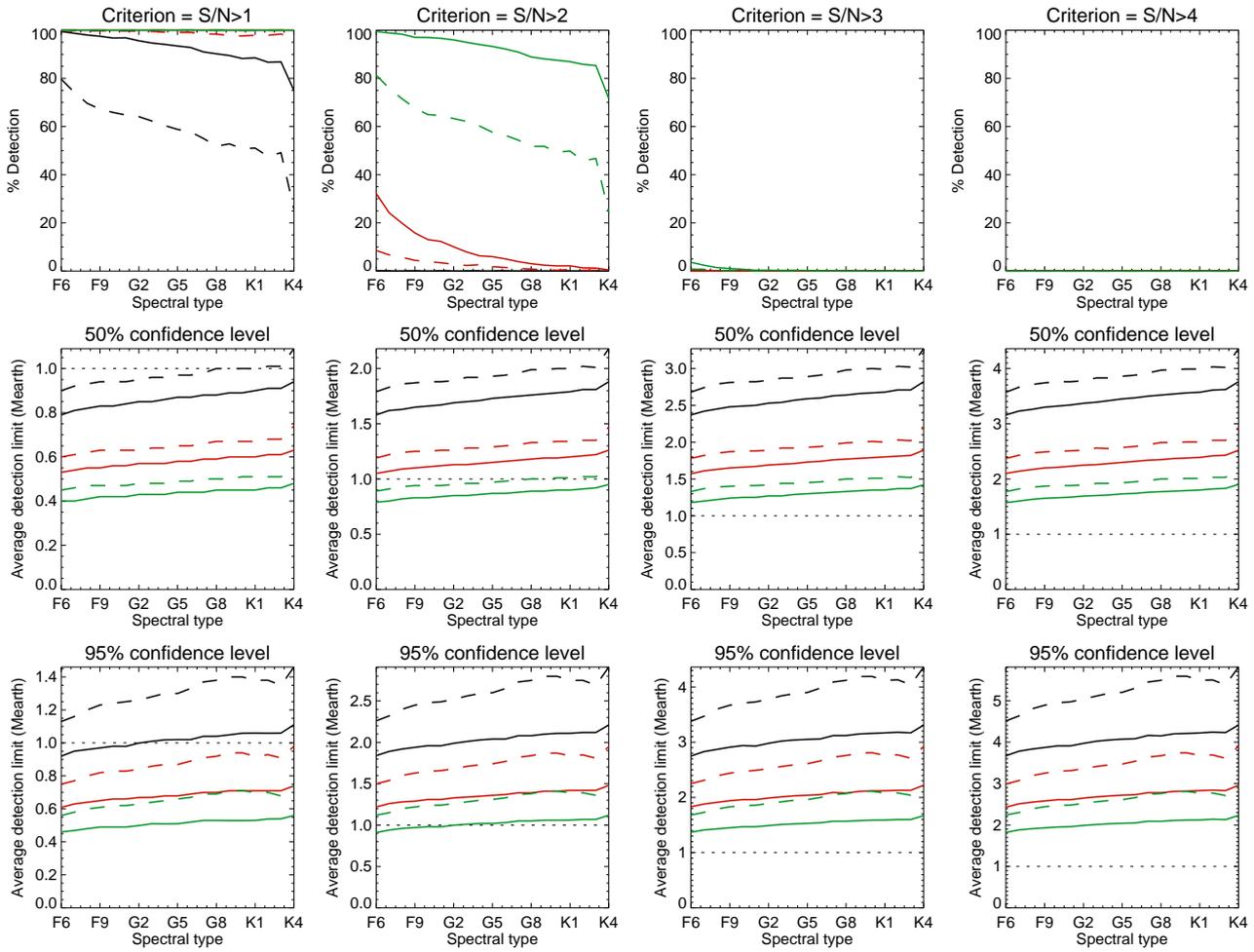}
\caption{
{\it First row:} Detection rate vs. spectral type deduced from different S/N thresholds (from left to right: 1, 2, 3, and 4) for $\Delta$T$_{\rm spot1}$ (solid) and $\Delta$T$_{\rm spot2}$ (dashed) for a 1 M$_{\rm Earth}$ planet at different orbital periods: lower HZ (black), medium HZ (red), and upper HZ (green).  Curves that are not visible are at the 100 \% level S/N thresholds 1 and 2, and at the 0 \% level for thresholds of 3 and 4.
{\it Second row:} Mass detection limit with a 50\% confidence level for the same criteria (same colour and line code). 
{\it Third row:} Same with a 95\% confidence level.
}
\label{sn}
\end{figure*}

 In this appendix,  we  compute detection rates based on a global S/N, as described in Sect.~2.4.1. 
 We considered S/N thresholds between 1 and 4. For a given S/N threshold, we computed the percentage of simulations with S/N above that threshold (for the 1 M$_{\rm Earth}$ planet), which provides the detection rate shown in the upper panel in Fig.~\ref{sn} for the three habitable zone orbital periods and the two spot contrasts. The detection rates are excellent when the threshold is equal to 1, except for the inner side of the habitable zone, where they can be as low as 20\% for K stars. They are much lower than 100\% for a a threshold of 2, with very poor performance for the inner habitable zone.  For higher thresholds (S/N of 3 and 4), the detection rate is equal to 0. This does not imply that a detection with S/N of 1 or 2 is a good detection, but we point out that with this standard definition of the S/N, a 1 M$_{\rm Earth}$ planet should not be detectable for stars like the Sun. We  discuss this S/N definition in  Sect.~4.2.5.  A much higher threshold \cite[such as 5 or 6 as used in some previous works, e.g.][]{theia17} would lead to a detection rate of 0\% as well for such low-mass planets, as $\alpha$ is  never higher than five or six times the noise for a 1 M$_{\rm Earth}$ planet (at 10 pc).  
We  compare these rates with more sophisticated approaches taking the temporal and frequential properties of the signal into account in Appendix G. 

The same approach can then be used to determine the planet mass corresponding to a certain detection rate (i.e. a certain percentage of simulations with an S/N above the considered threshold). The results are shown for 50\% and 95\% detection rates in Fig.~\ref{sn} (second and third row, respectively). Detection limits can be as low as 0.4  M$_{\rm Earth}$ for certain configurations (massive stars and $\Delta$T$_{\rm spot1}$) and are always below 3  M$_{\rm Earth}$ for the low thresholds, but they are between 2 and 4 M$_{\rm Earth}$ for thresholds of 3-4 on S/N.

\section{Comparison of FAP, fp, and LPA detection limits}

In this appendix, we discuss some  observational tools in more detail. We first compare the FAP with the theoretical false positive levels and then characterise the properties of the LPA detection limits. 

\subsection{Comparison of the FAP with the true false positive levels}

\begin{figure}
\includegraphics{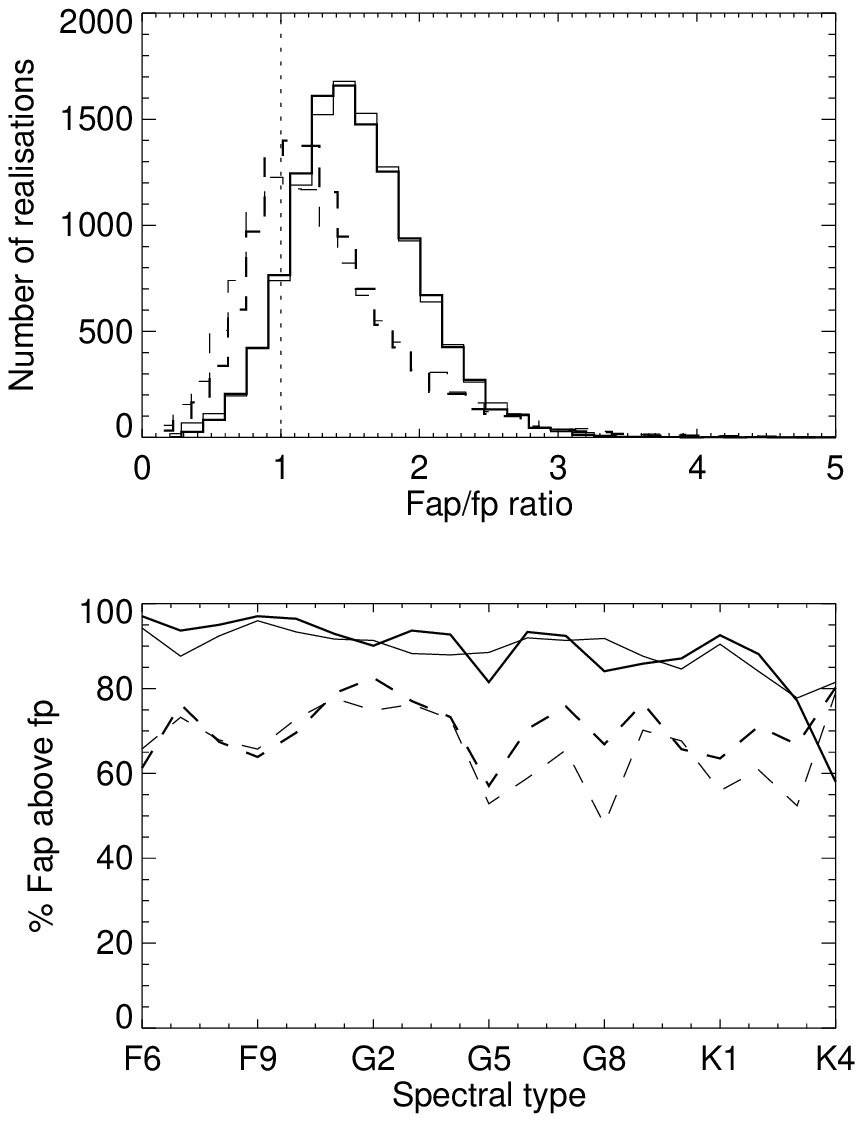}
\caption{
Distribution of the FAP/fp ratio (upper panel), and percentage of simulations for which the FAP is higher than the fp (lower panel) vs. spectral type. The solid lines correspond to $\Delta$T$_{\rm spot1}$ and the dashed lines to $\Delta$T$_{\rm spot2}$. The thin lines are for a constant fp and the thick lines for an fp dependent on stellar variability. }
\label{fig_fap}
\end{figure}

We compared the FAP and the theoretical fp  in a simplified configuration:  stars seen edge-on only, and no inclination between the orbital plane and the equatorial plane ($\Psi$=0). We  focused on the middle of the habitable zone. For each of these simulations (no planet was added), we computed the false positive level (as in Sect.~4.2) and the FAP level (using a bootstrap analysis with 1000 bootstrap iterations), both at the 1\% level. The FAP corresponds to the assumption that the signal is due to Gaussian noise. Our objective in this section is to compare these two estimates of the false positive levels, and the effect on the detection rates was studied in  the blind tests in Sect.~4.3. 
The ratio distributions of FAP/fp are shown  in Fig.~\ref{fig_fap}  (upper panel). There is an overlap with a ratio of 1, but the distributions are clearly shifted towards values higher than 1, that is, on average, the FAP overestimates the false positive level: this  should lead to a conservative approach when a detection is to be made.  
The ratio depends only slightly on spectral type, and the weak trend is not significant. The percentage of simulations for which the FAP is higher than fp lies between 60\% and 95\% depending on the spot contrast we considered.

\subsection{LPA detection limits}

\begin{figure}
\includegraphics{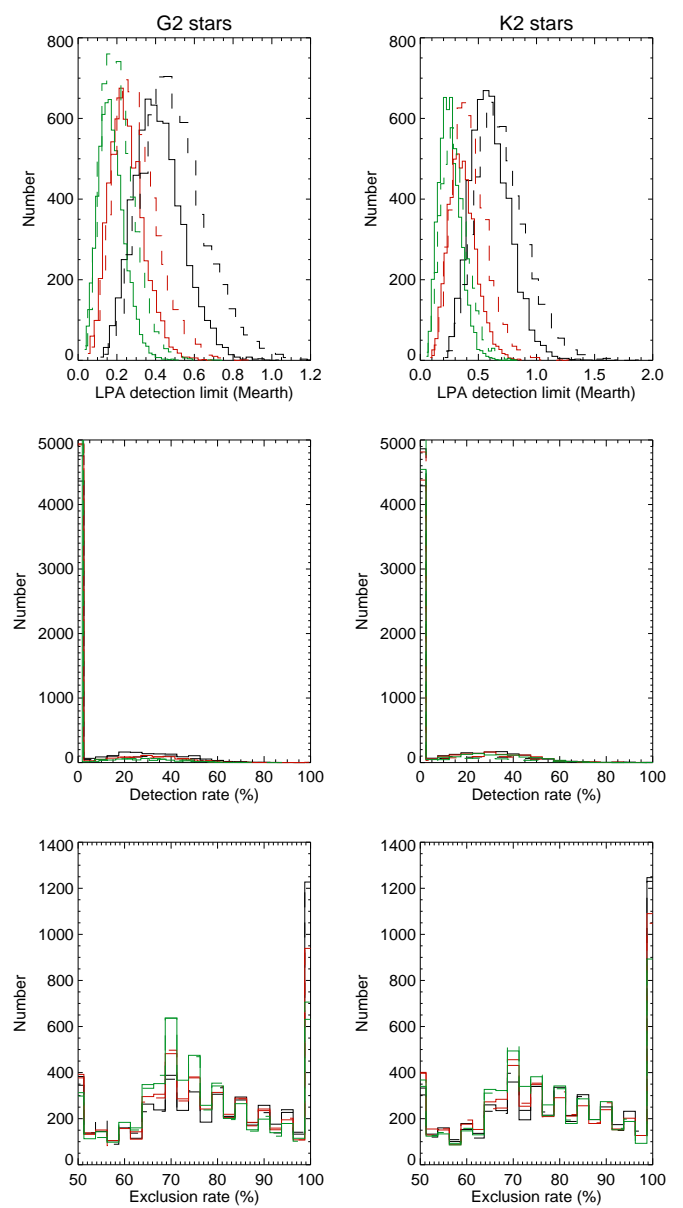}
\caption{
LPA caracterisation for G2 stars (left) and K2 stars (left). {\it First row:} Distribution of the LPA detection limits for $\Delta$T$_{\rm spot1}$ (solid lines) and $\Delta$T$_{\rm spot2}$ (dashed lines) for the different HZ periods: lower HZ (black), medium HZ (red), and upper HZ (green). {\it Third row:} Same for the distribution of the detection rates corresponding to the LPA masses. 
{\it Fourth row:} Same for the distribution of exclusion rates corresponding to the LPA masses. 
}
\label{lpa}
\end{figure}

We characterised the detection limits provided by the LPA approach \cite[][]{meunier12} in two cases, for G2 and K2 stars. The two spectral types lead to similar conclusions. Again, we focused on stars seen edge-on, without an inclination between the orbital plane and the equatorial plane. The principle of the LPA estimation, which is a fast method for computing detection limits, is as follows. The maximum of the periodogram around the period of interest was computed, and we searched for the planet mass that would produce a peak 1.3 times the observed peak amplitude \cite[][]{meunier12,lannier17}: we assumed that a more massive planet is excluded by the data because it would have produced a higher peak than observed. The obtained mass is therefore an exclusion rate, that is, we aim at excluding the presence of planets above a certain mass (the LPA detection limit) at the considered orbital period.

The distribution of detections limits found for the three values of orbital periods in the habitable zone and two spot contrasts are shown in Fig.~\ref{lpa} (upper panel): they cover a range between 0.1 and 1.5 M$_{\rm Earth}$, with distribution peaks at about 0.2 (for $\Delta$T$_{\rm spot1}$) and about 0.4-0.5 (for $\Delta$T$_{\rm spot2}$), that is, they are lower than the detection limits estimated in Sect.~4.2.4. This is explained by the fact that they correspond to a very low detection rate, as illustrated in the middle panels: for each simulation, we superposed 100 simulations of a planetary signal corresponding to the LPA mass (but different phases) and computed the periodogram and maximum power at the planet period. This allowed us to compute an exclusion rate for each simulation, as well as a detection rate by comparison with the theoretical fp. 
The distribution of exclusion rates is also shown in the lower panels. They are always higher than 50\%, with a strong peak at 100\%. 
For example, for G2 stars, the median exclusion rate is 78\%, and 17\% of the simulations have a 100\% exclusion rate.

\section{Additional blind test results}

In the appendix, we first show complementary results for the reference blind test A. Additional blind tests are then performed, in which individual parameters are modified compared to our reference blind test A (Table~\ref{tab_bt}). This allows us to show the effect of various conditions (stellar properties with C-F, observational strategy with N, instrumental noise with H, and planet properties with B, G, and I-M) on the detection rates and false positive rates. 
These blind tests were only made for one spectral type out of two, which is sufficient for the comparison. 
The average rates are shown in Table~\ref{tab_bt_all}. The false positive levels without a planet lie between 0.2 and 0.6\%, that is, they are relatively stable in the different tests, and lower than the 1\% FAP level used to perform the analysis. The false positive levels with a planet are usually also low, but depend on the planet parameter (mass and distance to its host star).

\begin{table}
\caption{List of blind tests}
\label{tab_bt}
\begin{center}
\renewcommand{\footnoterule}{}  
\begin{tabular}{lll}
\hline
Identification &  Properties \\
\hline
{\bf A} & {\bf reference blind test} \\
B & $\Psi$=0 \\
C & low $<\log R'_{HK}>$ (quiet) \\
D & high $<\log R'_{HK}>$  (active)\\
E & low Acyc (quiet) \\
F & high Acyc  (active)\\
G & 2 M$_{\rm Earth}$ \\
H & instrumental noise x 2 \\
I & star at 5 pc \\
J & star at 15 pc \\
K & star at 20 pc \\
L & 0.5 M$_{\rm Earth}$, star at 5 pc \\
M & 0.5 M$_{\rm Earth}$ \\
N1, N2, N3 & 3 tests on PHZ$_{\rm in}$, PHZ$_{\rm med}$, \& PHZ$_{\rm out}$ \\
O1, O2 & 100 observations (3.5 and 7 years)\\
\hline
\end{tabular}
\end{center}
\tablefoot{Properties are given by comparison with the reference blind test A (indicated in bold), which corresponds to 1 M$_{\rm Earth}$, star at 10 pc, all activity levels, instrumental noise of 0.199 $\mu$as per point, 50 observations over 3.5 years, distribution of $\Psi,$ and stellar inclination fixed to the true value.
}
\end{table}

\begin{table*}
\caption{Average rates for all blind tests}
\label{tab_bt_all}
\begin{center}
\renewcommand{\footnoterule}{}  
\begin{tabular}{c|cc|cccc}
\hline
Identification &  \multicolumn{2}{c|}{No injected planet} & \multicolumn{4}{c}{Injected planet} \\ \hline
  &  Good  & False  & Good  & Incorrect & Missed & Rejected \\
  & retrieval & positive & planet & planet & planet & planet \\
\hline
A & 99.6/99.6 & 0.4/0.4 & 92.5/80.0  & 0.5/0.7 & 0.4/3.2 & 6.6/16.2 \\
B & 99.6/99.6  & 0.4/0.4 & 91.9/79.7 & 0.5/0.6 & 0.9/3.9 & 6.7/15.8 \\
C & 99.8/99.6 & 0.2/0.4 & 92.8/83.4 & 0.6/0.8 & 0.7/3.1 & 5.9/12.7 \\
D &  99.7/99.6 & 0.3/0.4 & 91.0/76.4 & 0.6/1.0 & 0.6/3.8 & 7.8/18.8 \\
E & 99.8/99.8 & 0.2/0.2 & 92.4/85.4 & 0.4/0.4 & 0.6/2.6 & 6.5/11.6 \\
F & 99.5/99.7 & 0.5/0.3 & 91.4/75.8 & 0.4/0.7 & 0.7/4.1 & 7.7/19.3 \\
G & 99.6/99.6  & 0.4/0.4  & 99.8/99.8 & 0.2/0.2 & 0/0 & 0/0 \\
H & 99.7/99.8  &  0.3/0.2 &  27.6/21.9 & 1.1/0.7 & 29.0/34.6 & 42.4/42.7 \\
I & 99.6/99.6 & 0.5/0.4 & 99.8/99.7 & 0.2/0.2 & 0/0 & 0/0 \\
J & 99.8/99.8 & 0.2/0.2 & 54.2/35.9 & 1.3/1.2 & 12.3/25.3 & 31.6/37.6 \\
K & 99.7/99.8 & 0.3/0.2 & 23.0/12.5 & 1.4/1.1 & 36.4/53.0 & 39.2/33.5 \\
L & 99.7/99.8 & 0.3/0.2 & 91.7/79.8 & 0.5/0.7 & 0.8/4.0 & 7.1/15.5 \\
M & 99.4/99.8  & 0.6/0.2 & 30.1/18.6 & 2.2/1.6 & 31.6/45.8 & 36.1/34.1 \\N PHZ$_{\rm in}$  & 99.7/99.7 & 0.3/0.3 & 56.9/35.8 & 0.05/0.2 & 7.2/20.1 & 35.8/43.8 \\
N PHZ$_{\rm med}$ & 99.6/99.8 & 0.4/0.2 & 96.0/84.2 & 0/0.05 & 0.1/1.3 & 3.9/14.5 \\
N PHZ$_{\rm out}$ & 99.6/99.7 & 0.4/0.3 & 97.7/95.0 & 2.2/2.8 & 0/0  & 0.1/2.2 \\
O (3.5 y) & 97.4/97.4 & 2.6/1.6 & 99.6/98.6 & 0.4/0.5 & 0/0.1 & 0/0.8 \\
O (7 y) & 97.9/98.1 & 2.1/1.9 & 99.9/99.0 & 0.04/0.1 & 0/0.2 & 0.05/0.6 \\
\hline
\end{tabular}
\end{center}
\tablefoot{Percentages are averaged over all spectral types and are given for $\Delta$T$_{\rm spot1}$ and $\Delta$T$_{\rm spot2}$ (separated by the slash). They correspond to an FAP level of 1\%.  The properties of each of these blind tests can be found in Table~\ref{tab_bt}. 
}
\end{table*}

\subsection{Dependence on spectral type for reference blind test A}

\begin{figure}
\includegraphics{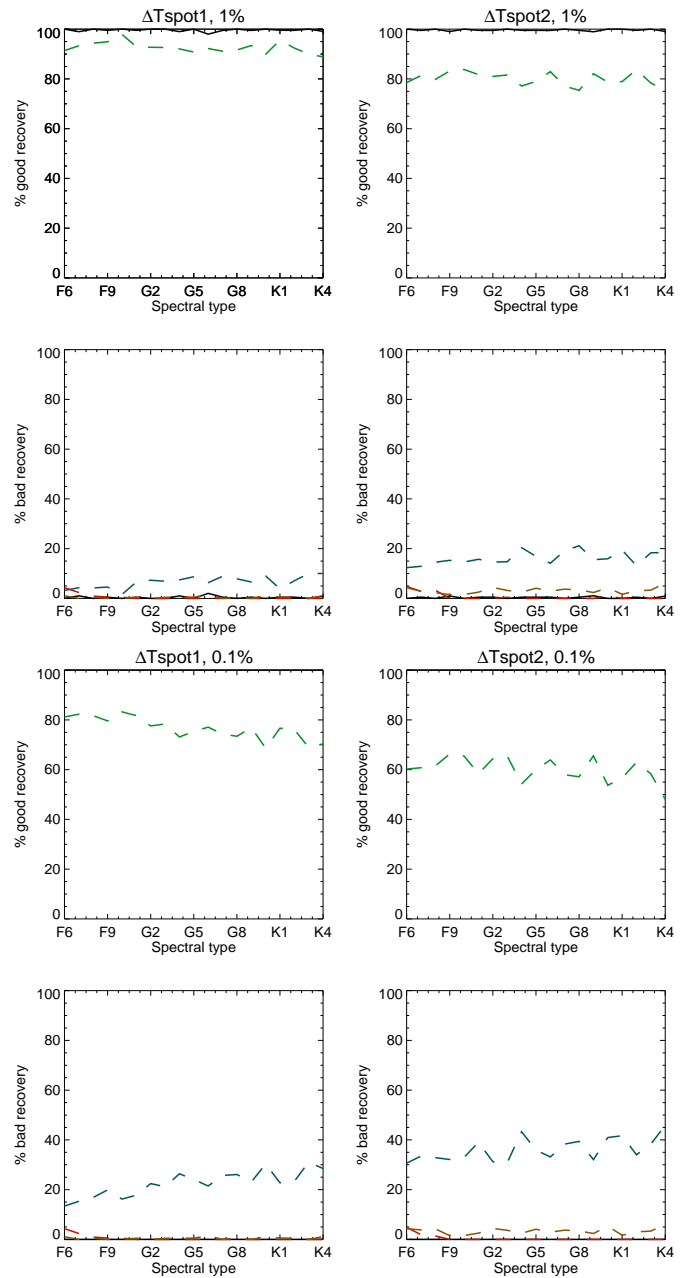}
\caption{
Percentage of good and poor recovery rates vs. spectral type for 1\% FAP level (two first rows) and 0.1\% FAP level (two last rows) from blind tests A. The good recovery rates correspond to the case without a planet (solid black line) and with a planet (dashed green line). The poor recovery rates correspond to the case without a planet (solid black line, false positive) and with a planet: incorrect planets (red dashed line, false positive), rejected planets (blue dashed line), and missed planets (brown dashed line). In the third row all curves except one are at the 100\% level, and most poor recovery curves are close to 0\%. }
\label{btA_taux}
\end{figure}

 Figure~\ref{btA_taux} shows the detection rates and the various rates of poor recovery  versus spectral type for the reference blind test A. They are discussed in Sect.~4.2.4. The average rates are shown for all blind tests in Table~\ref{tab_taux_moyen}.

\subsection{Effect of orbit inclination with respect to the line of sight (blind test A)}

We also compared the detection rates for various inclinations of the orbit with respect to the line of sight in the case of blind test A. For this purpose, we  computed the ratio $r$ between the apparent semi-major axis and the apparent semi-minor axis (true parameters) for all injected   planets. The simulations were then separated into two subsets depending on the value of $r$: values of $r$ close to one correspond to close to circular orbit (orbit seen pole-on), and high values correspond to orbits close to edge-on. Table~\ref{tab_taux_orbit} shows the resulting detection rates for various selections. We conclude that the planets with a pole-on configuration are better detected than the other configurations, as already seen for more simple noise contribution by \cite{eisner01}: this is also true here despite the complex  behaviour of the activity signal in the two directions.

\begin{table}
\caption{Detection rates (in percent) depending on planet orbit inclination with respect to the line of sight.}
\label{tab_taux_orbit}
\begin{center}
\renewcommand{\footnoterule}{}  
\begin{tabular}{lll}
\hline
 & $\Delta$T$_{\rm spot1}$ & $\Delta$T$_{\rm spot2}$ \\ \hline
$r<$1.25 & 99.1& 95.1\\
$r>$5 & 83.3& 64.6\\
$r<$med($r$) & 97.1& 89.6\\
$r>$med($r$) & 87.9& 70.4\\
\hline
\end{tabular}
\end{center}
\tablefoot{The ratio $r$ is computed between  the apparent semi-major axis and the apparent semi-minor axis of the  injected planet.
}
\end{table}

\subsection{Results for blind test B: $\Psi$=0}

The average rates for blind test B, that is, $\Psi$=0, to compare with reference blind test A, are shown in Table~\ref{tab_bt_all}. The rates from blind tests A and B are extremely similar, which means that the exact distribution of the angle between orbital plane and equatorial plane $\Psi$ is not critical and does not affect our results. For this reason,  all other blind tests were  made with the distribution of $\Psi$  used in blind test A.

\subsection{Results for blind tests C and D: low and high average activity level}

The average rates for blind tests C and D, that is, for low $<\log R'_{HK}>$ (quiet stars) 
and  high $<\log R'_{HK}>$  (active stars), respectively, to compare with reference blind test A, are shown in Table~\ref{tab_bt_all}. For $\Delta$T$_{\rm spot1}$, the effect is very weak, with very similar (and high) detection rates for quiet and active stars.
On the other hand, for  $\Delta$T$_{\rm spot2}$, the detection rate with planet is slightly better for quiet stars (83\% instead of 76\%), although the difference is not great.

\subsection{Results for blind tests E and F: low and high Acyc (quiet)}

The average rates for blind tests E and F, that is, for low cycle amplitude and  high cycle amplitude, respectively, to compare with reference blind test A, are shown in Table~\ref{tab_bt_all}. The effect is similar to the selection based on the average activity level (blind tests C and D), with a better detection rate for quiet stars.

\subsection{Results for blind test G: 2 M$_{\rm Earth}$ }

The average rates  for blind test G, that is, a 2 M$_{\rm Earth}$ planet, to compare with reference blind test A, are shown in Table~\ref{tab_bt_all}. The performance is excellent in this case, with almost no false positive and no missed planets. 
 
\subsection{Results for blind test H: instrumental noise x 2}

The average rates for blind test H, that is, an instrumental noise twice higher, to compare with reference blind test A, are shown in Table~\ref{tab_bt_all}. The effect is not significant when no planet is injected, but the detection rates are very sensitive to this parameter. 

\subsection{Results for blind tests I, J, and K: star at 5 pc, 15pc, and 20 pc}

The average rates  for blind tests I, J, and K, that is, different stellar distances, to compare with reference blind test A, are shown in Table~\ref{tab_bt_all}. Stellar distance significantly affects the detectability, as expected. At 15 pc, the rates decrease to 54-36\% depending on the spot contrast that is assumed, with many rejected planets. They are below 23\% for a distance of 20 pc. 

\subsection{Results for blind tests L and M: 0.5 M$_{\rm Earth}$, star at 5 pc and 10 pc}

The average rates for blind tests L and M, that is, a  0.5 M$_{\rm Earth}$ planet, to compare with reference blind test A, shown in Table~\ref{tab_bt_all}. The detection rates for a  0.5 M$_{\rm Earth}$ planet at 5 pc is very similar to the rates for a  1 M$_{\rm Earth}$ at 10 pc (blind test A).  If the  0.5 M$_{\rm Earth}$ planet is at 10 pc, the detection rates are naturally much lower, but they show that some of these planets could be detected with astrometry.

\subsection{Results for blind tests N: PHZ$_{\rm in}$, PHZ$_{\rm med}$, and  PHZ$_{\rm out}$}

The average rates for the three blind tests N, that is, with planet orbital periods fixed specifically at PHZ$_{\rm in}$, PHZ$_{\rm med}$, and PHZ$_{\rm out}$ in order to compare with the computations made in Sect.~4.2 with the theoretical false positive levels, are shown in Table~\ref{tab_bt_all}. The results are discussed in Sect.~4.5. 

\subsection{Results for blind tests O: 100 observations}

We performed  two blind tests O, that is, with a 100 sampling (instead of 50), covering either 3.5 years as blind test A, or seven years.
Our main objective was to determine whether a higher number of points can improve the significance of the planet peaks that were below the FAP (but at the proper period) in blind test A, as well as the missed planets for the most active stars. We considered two strategies. In the first (blind test O1), we observed 100 points over the same duration as before, which could correspond to a situation where the star has been identified as active (e.g. from previous photometric data), and which may require more points to reach a good performance. The second strategy (blind tests O2), with 100 points covering seven years, could correspond to stars with no detection during the first 3.5 years but the presence of a peak above the noise level (with a peak S/N defined as in Sect.~4.3.5) although below the FAP, high enough to suggest the possible presence of a planet (candidate), requiring additional data. With 100 points, the detection rates of injected planets are very close to 100\%, which is better than the 50-point configuration. The average rates  are shown in Table~\ref{tab_bt_all}. There is not strong difference between the 3.5 y and 7 y coverages. We note that the false positive level without a planet is slightly higher than for 50 points given a similar criterion on the FAP level, so that we caution about this side effect. It may be mitigated, for example, by also considering subsets of data.

\section{Comparison of detection rates using the different approaches}

\begin{figure}
\includegraphics{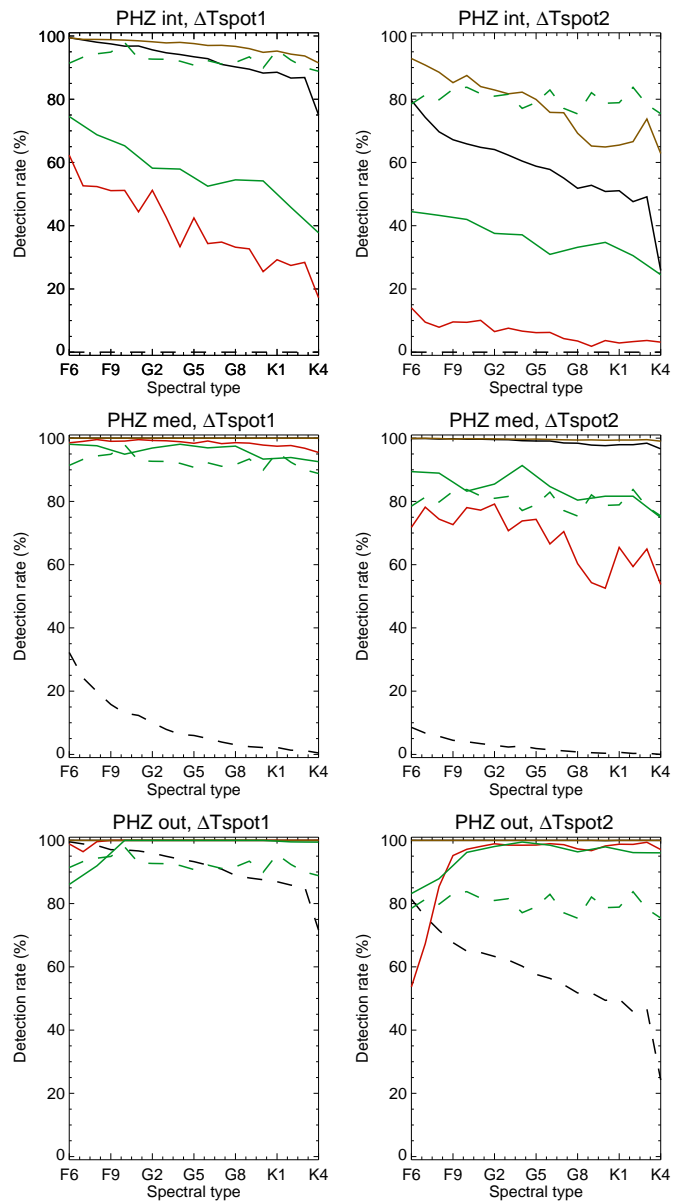}
\caption{
Comparison of detection rates vs. spectral type for different $\Delta$T$_{\rm spot}$ and orbital periods in the habitable zone for different conditions: based on S/N$>$1 (solid black lines), on S/N$>$2 (dashed black lines), true 1\% fp from frequential analysis (brown lines), true 1\% fp from temporal analysis (red lines), and blind test A (dashed green line, covering the whole habitable zone), and blind tests N (green lines). 
 }
\label{recap}
\end{figure}

In this  section, we compare the detection rates obtained in previous sections using different approaches for a 1 M$_{\rm Earth}$ planet. The detection rates versus spectral types are shown in Figure~\ref{recap}. We first compare the rates obtained with the S/N threshold and the theoretical fp. The two correspond reasonably well for the S/N$>$1 computation, while S/N$>$2 leads to very poor rates compared to what could be theoretically be achieved. The agreement is best for fp based on power (frequential analysis). For PHZ$_{\rm in}$, $\Delta$T$_{\rm spot2}$, the rates from fp are slightly better than those obtained with S/N$>$1. 

Figure~\ref{recap} also allows us to compare the detection rates obtained with the true fp (frequential analysis) with the blind tests. Because blind test A was implemented for the whole habitable zone (dashed lines), that is, the periods were randomly chosen in the whole habitable zone, we also performed blind tests dedicated to the three orbital periods (blind tests N1, N2, and N3, see Appendix F.10) for a proper comparison. The trends are similar, but for PHZ$_{\rm in}$ and PHZ$_{\rm med}$, for which the detection rates are not as close to 100\% as for  PHZ$_{\rm out}$, the detection rates are much lower from the blind test compared to the use of the true fp  level (frequential analysis). The reason is that the FAP evaluation in the blind test is overestimated compared to fp (and particularly when a planet is present), meaning that the effective false positive level in the blind test is much lower than 1\% and some planets are undetected even though the highest peak is at the true planet period. This leads to a significant difference between the two estimates. 

Finally, we underline that the performance is very sensitive to the position inside the habitable zone. The performance is excellent in the outer part of the habitable zone, but significantly poorer (for the distance and planet mass considered here) in the inner part.

\end{appendix}

\end{document}